\documentclass{article}
\usepackage{amssymb,hyperref}
\usepackage[english]{babel}
\usepackage{amsmath, graphics}
\usepackage{geometry}
\usepackage{color}
\newcommand{\mathsym}[1]{{}}
\newcommand{\unicode}[1]{{}}

\def\noi{\noindent}

\def\nqq{\hspace*{-2em}}

\def\lal{&&\nqq {}}
\def\nnn{\nonumber\\ \lal }
\def\nl{\\ \lal }
\def\nll{\nonumber \nl\hspace{2.2em} }

\def\barr{\left(\begin{array}}
\def\earr{\end{array}\right)}
\def\beq#1{\begin{equation}\label{#1}}
\def\eeq{\end{equation}}
\def\ber#1{\begin{eqnarray}\label{#1} &&\nqq}
\def\eer{\end{eqnarray}}

\newcommand{\bear}[1]{\begin{eqnarray}\label{#1}}
\newcommand{\bearr}[1]{\begin{eqnarray}\lal \label{#1}}
\newcommand{\ear}{\end{eqnarray}}

\catcode`\@=11 \@addtoreset{equation}{section}\catcode`\@=12

\newcommand{\R}{ {\mathbb R} }

\newcommand{\fnm}{\footnotemark}
\newcommand{\fnt}{\footnotetext}

\begin{document}

 \vspace{15pt}

 \begin{center}
 \large\bf

 On generalized Melvin solutions for Lie algebras of rank 4

 \vspace{15pt}

 \normalsize\bf
      S. V. Bolokhov\fnm[1]\fnt[1]{bol-rgs@yandex.ru}$^{,a}$
      and  V. D. Ivashchuk\fnm[2]\fnt[2]{ivashchuk@mail.ru}$^{,a,b}$

 \vspace{7pt}

 \it 
 
 (a) 
  \ Peoples' Friendship University of Russia
   (RUDN University), \\ 
   6 Miklukho-Maklaya Street,
   Moscow, 117198, Russian Federation \\
 (b) \ \ \ Center for Gravitation and Fundamental Metrology, \\
  VNIIMS, 46 Ozyornaya St., Moscow 119361, Russian Federation \\

 \end{center}
 \vspace{15pt}

 \small\noi

 \begin{abstract}
 We deal with generalized Melvin-like solutions associated with Lie algebras of rank $4$ ($A_4$, $B_4$, $C_4$, $D_4$, $F_4$). Any solution has static cylindrically-symmetric metric in $D$ dimensions in presence of four Abelian 2-forms and four scalar fields. The solution is governed by four moduli functions $H_s(z)$ ($s = 1,...,4$) of squared radial coordinate $z=\rho^2$ obeying four differential equations of the Toda chain type. These functions are polynomials of powers $(n_1,n_2, n_3, n_4) = (4,6,6,4), (8,14,18,10), (7,12,15,16), (6,10,6,6), (22,42,30,16)$ for Lie algebras $A_4$, $B_4$, $C_4$, $D_4$, $F_4$, respectively. The asymptotic behaviour for the polynomials at large $z$ is governed by an integer-valued $4 \times 4$ matrix $\nu$ connected in a certain way with the inverse Cartan matrix of the Lie algebra and (in $A_4$ case) the matrix representing a generator of the $\mathbb{Z}_2$-group of symmetry of the Dynkin diagram. The symmetry properties and duality identities for polynomials are studied. We also present 2-form flux integrals over a $2$-dimensional submanifold.  Dilatonic black hole analogs of the obtained Melvin-type solutions, e.g. ``phantom'' ones, are also considered. The phantom black holes are described by fluxbrane polynomials under consideration.
  
  \end{abstract}

Key-words: Melvin solution; fluxbrane polynomials; Lie algebras; dilatonic black hole, phantom black holes 

\large 

 \section{Introduction}

  In this  semi-review paper, we study  multidimensional generalization of Melvin's solution \cite{Melv}, which was presented earlier in ref. \cite{GI-09}. Originally, model from ref. \cite{GI-09}
  contains metric, $n$ Abelian 2-forms and  $l \geq n$ scalar fields. Here we consider a special solutions with $n =l =4$, governed by a $4 \times 4$ Cartan matrix  
  $(A_{i j})$ for Lie algebras of rank $4$: $A_4$, $B_4$,  $C_4$, $D_4$, and the exceptional algebra $F_4$. The solutions from ref. \cite{GI-09} are special case of the so-called generalized fluxbrane solutions from ref. \cite{Iflux}.  
  
   The original Melvin's  $4d$ solution  describes the gravitational field of a magnetic flux tube. The multidimensional analog of such a flux tube, supported by a certain configuration of form fields, 
   is   referred to as a fluxbrane. Earlier the appearance of fluxbrane 
  solutions was related mainly to supergravity models with motivations supported by    
    by superstring/$M$-theory approach. For generalizations of the Melvin solution and  fluxbrane solutions see  
  \cite{BronShikin}-\cite{Ivas-Symmetry-17} and references therein.  

  In ref. \cite{Iflux} there were considered the generalized fluxbrane solutions which are described in terms of moduli functions $H_s(z) > 0$ defined on the interval $(0, +\infty)$, where $z = \rho^2$ and $\rho$ is a radial coordinate. Functions $H_s(z)$ obey $n$ non-linear differential master equations of Toda-like type governed by some matrix $(A_{s s'})$, and the following boundary conditions are imposed: $H_{s}(+ 0) = 1$,   $s = 1,...,n$.  

  Here  we put the matrix  $(A_{s s'})$ to be coinciding with a Cartan matrix for some 
 simple finite-dimensional Lie algebra $\cal G$ of rank $n$. 
 It was  conjectured in ref. \cite{Iflux} that in this case 
 the solutions to master equations with the above boundary conditions 
 are  polynomials of the form: 
  \beq{1.3}
   H_{s}(z) = 1 + \sum_{k = 1}^{n_s} P_s^{(k)} z^k,
  \eeq
  where $P_s^{(k)}$ are constants. Here
 $P_s^{(n_s)} \neq 0$  and 
 \beq{1.4}
  n_s = 2 \sum_{s' =1}^{n} A^{s s'},
 \eeq 
 where we denote $(A^{s s'}) = (A_{s s'})^{-1}$.
 Integers $n_s$ are components  of the twice dual
 Weyl vector in the basis of simple (co-)roots \cite{FS}.
 
For any simple finite-dimensional Lie algebra $\cal G$ 
the functions $H_s$, which are called``fluxbrane polynomials'', 
define a special solution to open Toda chain equations \cite{K,OP} corresponding to $\cal G$ \cite{I-14}.
It was pointed out in ref. \cite{Iflux} that the conjecture on polynomial structure of  $H_{s}(z)$ is valid for 
all Lie algebras of $A$- and $C$- series.

Here we study some geometric properties of the solutions  corresponding to Lie algebras of rank $4$: 
we present some symmetry relations and duality identities of fluxbrane polynomials. The latter are controlling the
 transformations $\rho \to 1/\rho$  and depend upon  the groups of symmetry 
   of Dynkin diagrams for  Lie algebras. In our case these groups of symmetry are trivial
     (i.e. identical) ones  for Lie algebras $B_4$, $C_4$ and $F_4$, while for the Lie algebra $A_4$
    we get the group $\mathbb{Z}_2$ and for the Lie algebra $D_4$ we are led to symmetric group $S_3$.  
      
   The analogous analysis was done earlier for the case of rank-2 Lie algebras: 
   $A_2$, $B_2 = C_2$, $G_2$  in ref. \cite{BolIvas-R2-17}, and for rank-3 algebras $A_3$, $B_3$, $C_3$ in Ref. \cite{BolIvas-R3-18}. Also, in ref. \cite{BolIvas-17}  the conjecture from ref. \cite{Iflux} was verified for the Lie algebra $E_6$ and certain duality relations for six $E_6$-polynomials were found.

The paper is organized as follows.
In Section 2 we present a generalized  Melvin solutions
from ref. \cite{GI-09} for the case of four scalar fields and four 2-forms. In Section 3 we deal with the solutions for the Lie algebras $A_4$,  $B_4$,  $C_4$, $D_4$ \cite{Bol-Ivas-R4-18} and  $F_4$.
 We present symmetry properties,  duality relations for polynomials and 2-form flux integrals $\Phi^s = \int F^s$ over a $2d$ submanifold, where  $F^s$ are 2-forms   \cite{Ivas-flux-17}. In Section 4 we consider black hole analogs of the obtained Melvin-type solutions, e.g. phantom ones.

It should be noted that the fluxbrane polynomials, which give us special solutions to Toda chain equations, may be useful  for describing  supergravity model solutions. Indeed, let us restrict ourselves to  maximal supergravity models in dimensions $D < 11$  \cite{LP0} which are obtained from $D= 11$ supergravity by dimensional reductions on tori. It was shown in ref. \cite{LMMP} that there exist special cosmological  and  static cylindrically-symmetric domain wall solutions in dimensions $D= 3, 4, 5,6,7$, which are described by Toda equations corresponding to $E_N$ Lie algebras with $N = 11 - D$, where $E_6, E_7, E_8$ are standard exceptional Lie algebras and  $E_5 = D_5$, $E_4 = A_4$. 
\fnm{} \fnt{In ref. \cite{LMMP} the existence of polynomial Toda chain solutions corresponding to 
$E_8$ Lie algebra (with proper powers of polynomials) was conjected and  polynomials  related to $D_4$ Lie algebra were presented. }  By putting a certain charge (corresponding to off-line root in Dynkin diagram) to zero  we get $A_{N-1}$ Toda chains (TC) ($N= 4,5,6,7,8$)  \cite{LP1},  while identifying certain pairs of charges we get $F_4$ TC from $E_6$ one,  $B_4$ TC from $D_5$ one and  $C_4$ TC from $A_7$ one, see ref.  \cite{GI}. The $D_5$ solution with a certain charge equal to zero gives us a $D_4$ solution. For  TC solutions (e.g. black brane and fluxbrane ones) in supergravitational models corresponding to Lie algebras of lower ranks (e.g. $A_1$, $A_2$)  see  \cite{Ivas-Symmetry-17,DLP,LPX,IMtop} and references therein.

Another possible application of the results of this and previous our works on fluxbrane polynomials may be in considering of obtained $4d$ dilatonic solutions as  backgrounds for studying of  so-called quasinormal modes  \cite{KonZh}  and related problems (photon spheres, shadows, echoes etc). This topic is rather popular at present, especially after the discovery of gravitational waves.

\section{The set up and generalized Melvin solutions}

Let us consider the following product manifold:
\beq{2.2}
  M = (0, + \infty)  \times M_1 \times M_2,
 \eeq
 where $M_1 = S^1$ and
 $M_2$ is a $(D-2)$-dimensional Ricci-flat manifold.

 We define the action 
 \beq{2.1}
 S =\int_{M} d^Dx \sqrt{|g|} \biggl \{R[g]-
 \delta_{a b} g^{MN}\partial_M \varphi^{a} \partial_N \varphi^{b} - \frac{1}{2}
 \sum_{s =1}^{4}\exp[2 \vec{\lambda}_s \vec{\varphi}](F^s)^2 \biggr \},
 \eeq
 where $g=g_{MN}(x)dx^M\otimes dx^N$ is a metric on $M$,
 $\vec{\varphi} = (\varphi^a)\in \R^4$ is vector of scalar fields,
   $ F^s =    dA^s
          =  \frac{1}{2} F^s_{M N}  dx^{M} \wedge  dx^{N}$
 is a $2$-form,  $\vec{\lambda}_s = (\lambda_{s}^{a}) \in \R^4$ is dilatonic  coupling vector,
   $s = 1,...,4$; $a =1,...,4$.
 Here we use the notations $|g| \equiv  |\det (g_{MN})|$, $(F^s)^2 \equiv 
   F^s_{M_1 M_{2}} F^s_{N_1 N_{2}}  g^{M_1 N_1} g^{M_{2} N_{2}}$.

We deal with a family of exact
cylindrically-symmetric solutions to the field equations corresponding for the action
(\ref{2.1}) and depending on the radial coordinate $\rho$. 
  The solution has the form \cite{GI-09}:
 \bear{2.30}
  g= \Bigl(\prod_{s = 1}^{4} H_s^{2 h_s /(D-2)} \Bigr)
  \biggl\{  d\rho \otimes d \rho  +
  \Bigl(\prod_{s = 1}^{4} H_s^{-2 h_s} \Bigr) \rho^2 d\phi \otimes d\phi +
    g^2  \biggr\},
 \\  \label{2.31}
  \exp(\varphi^a)=
  \prod_{s = 1}^{4} H_s^{h_s  \lambda_{s}^a},
 \\  \label{2.32a}
  F^s =  q_s \left( \prod_{l = 1}^{4}  H_{l}^{- A_{s
  l}} \right) \rho d\rho \wedge d \phi,
  \ear
 $s, a = 1, ... , 4$, where  $g^1 = d\phi \otimes d\phi$ is a
  metric on $M_1 = S^1$ and $g^2$ is a  Ricci-flat metric of 
  signatute $(-,+, \dots, +)$ on
 $M_{2}$.  Here $q_s \neq 0$ are integration constants  
 ($q_s = - Q_s$ in notations of ref. \cite{GI-09}).

 Here we denote $z = \rho^2$. As it was shown in earlier works, the functions $H_s(z) > 0$ obey the set of master equations
\beq{1.1}
  \frac{d}{dz} \left( \frac{ z}{H_s} \frac{d}{dz} H_s \right) =
   P_s \prod_{l = 1}^{4}  H_{l}^{- A_{s l}},
  \eeq
 with  the boundary conditions
 \beq{1.2}
   H_{s}(+ 0) = 1,
 \eeq
 where
 \beq{2.21}
  P_s =  \frac{1}{4} K_s q_s^2,
 \eeq
 $s = 1,..., 4$.  The boundary condition (\ref{1.2}) guarantees the absence of a conic singularity (for the metric  (\ref{2.30})) for $\rho =  +0$.
 
 There are some relations for the parameters $h_s$:
  \beq{2.16}
  h_s = K_s^{-1}, \qquad  K_s = B_{s s} > 0,
  \eeq
 where
 \beq{2.17}
  B_{s l} \equiv
  1 +\frac{1}{2-D}+  \vec{\lambda}_{s} \vec{\lambda}_{l} ,
  \eeq
 $s, l = 1,..., 4$.
 In these relations, we have denoted 
 \beq{2.18}
  (A_{s l}) = \left( 2 B_{s l}/B_{l l} \right).
 \eeq
 The latter matrix is the so-called ``quasi-Cartan'' matrix. 
  One can prove that if $(A_{s l})$ is a Cartan matrix 
 for a certain simple Lie algebra $\cal G$ of rank $4$
 then there exists a set of vectors 
 $\vec{\lambda}_1, ..., \vec{\lambda}_4$  obeying (\ref{2.18}). See also Remark 1 in the next section.

The solution under consideration can be understood as a special case of the
fluxbrane solutions from \cite{Iflux,GIM}.

Therefore, here we investigate a multidimensional 
generalization of Melvin's solution \cite{Melv} for the case of four scalar fields and four $2$-forms. Note that the original Melvin's solution without scalar field would correspond to $D = 4$, one (electromagnetic) $2$-form, $M_1 = S^1$ ($0 < \phi <  2 \pi$),  $M_2 = \R^2$ and $g^2 = -  dt \otimes dt + d x \otimes d x$.

\section{Solutions related to  simple classical rank-4 Lie algebras}

In this section we consider the solutions associated with Lie algebras ${\cal G}$ of rank $4$. This means than the matrix  $A = (A_{sl})$ coincides with one of the Cartan matrices 
\begin{align} 
     \left(A_{ss'}\right) &=
     \begin{pmatrix}
     2 & -1 & 0 & 0 \\
     -1 & 2 & -1 & 0 \\
     0 & -1 & 2 & -1 \\
     0 & 0 & -1 & 2 \\
     \end{pmatrix},\!
  \quad
   \begin{pmatrix}
    2 & -1 & 0 & 0 \\
    -1 & 2 & -1 & 0 \\
    0 & -1 & 2 & -2 \\
    0 & 0 & -1 & 2 \\
  \end{pmatrix},\! 
  \quad
  \begin{pmatrix}
     2 & -1 & 0 & 0 \\
     -1 & 2 & -1 & 0 \\
     0 & -1 & 2 & -1 \\
     0 & 0 & -2 & 2 \\
    \end{pmatrix}, 
   \quad \nonumber \\
&\hspace{6em}   \begin{pmatrix}
        2 & -1 & 0 & 0 \\
        -1 & 2 & -1 & -1 \\
        0 & -1 & 2 & 0 \\
        0 & -1 & 0 & 2 \\
    \end{pmatrix},
     \begin{pmatrix}
            2 & -1 & 0 & 0 \\
            -1 & 2 & -2 & 0 \\
            0 & -1 & 2 & -1 \\
            0 &0 & -1 & 2 \\
        \end{pmatrix}
  \label{A.5}
     \end{align}
for ${\cal G} = A_4, \, B_4, \, C_4, \, D_4, \, F_4$, respectively. 

Each of these matrices can be graphically described by  Dynkin diagrams pictured on Fig. 1.
 \vspace{25pt}
  
  \setlength{\unitlength}{1mm}
 \begin{figure}[h]
 \centering
 \begin{picture}(140, 25)
 \put(2,25){\circle*{2}}
 \put(12,25){\circle*{2}}
 \put(22,25){\circle*{2}}
 \put(32,25){\circle*{2}}
 \put(1,25){\line(1,0){30}}
 \put(0.3,20.5){$1$}
 \put(10.3,20.5){$2$}
 \put(20.3,20.5){$3$}
 \put(30.3,20.5){$4$}
  
 \put(42,25){\circle*{2}}
 \put(52,25){\circle*{2}}
 \put(62,25){\circle*{2}}
 \put(72,25){\circle*{2}}
 \put(42,25){\line(1,0){20}}
 \put(62,25.5){\line(1,0){10}}
 \put(62,24.5){\line(1,0){10}}
 \put(40.3,20.5){$1$}
 \put(50.3,20.5){$2$}
 \put(60.3,20.5){$3$}
 \put(70.3,20.5){$4$}
 \put(65.3,23.9){\large $>$}
 
 \put(82,25){\circle*{2}}
 \put(92,25){\circle*{2}}
 \put(102,25){\circle*{2}}
 \put(112,25){\circle*{2}}
 \put(82,25){\line(1,0){20}}
 \put(102,25.5){\line(1,0){10}}
 \put(102,24.5){\line(1,0){10}}
 \put(80.3,20.5){$1$}
 \put(90.3,20.5){$2$}
 \put(100.3,20.5){$3$}
 \put(110.3,20.5){$4$}
 \put(105.3,23.9){\large $<$}
  
 \put(122,25){\circle*{2}}
 \put(132,25){\circle*{2}}
 \put(139,32){\circle*{2}}
 \put(139,18){\circle*{2}}
 \put(121,25){\line(1,0){10}}
 \put(131.3,24.5){\line(1,1){7.5}}
 \put(132,25){\line(1,-1){7.5}}
 \put(120.3,20.5){$1$}
 \put(130.3,20.5){$2$}
 \put(141,31){$3$}
 \put(141,16.3){$4$}

 \put(52,12){\circle*{2}}
 \put(62,12){\circle*{2}}
 \put(72,12){\circle*{2}}
 \put(82,12){\circle*{2}}
 \put(52,12){\line(1,0){10}}
 \put(72,12){\line(1,0){10}}
 \put(62,12.5){\line(1,0){10}}
 \put(62,11.5){\line(1,0){10}}
 \put(50.3,7.5){$1$}
 \put(60.3,7.5){$2$}
 \put(70.3,7.5){$3$}
 \put(80.3,7.5){$4$}
 \put(65.3,10.9){\large $>$}
 
  \end{picture}
  \vspace{-15pt}
  \caption{The Dynkin diagrams for the Lie algebras $A_4$, $B_4$, $C_4$, $D_4$, $F_4$, respectively.}
 \end{figure}
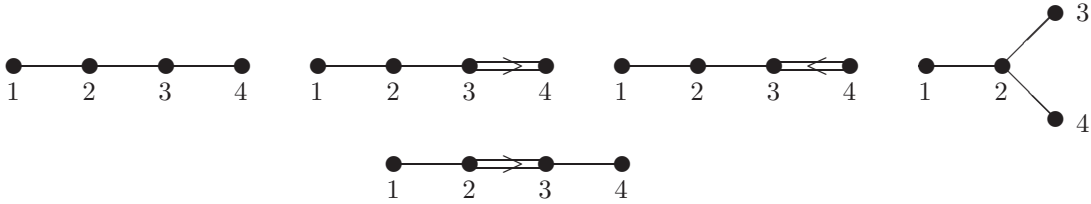
     
  Using (\ref{2.16})-(\ref{2.18}) we can get
    \begin{equation}
           \label{3.17}
    K_s =   \frac{D - 3}{D -2} +  \vec{\lambda}_{s}^2,
      \end{equation} 
    where $h_s =  K_s^{-1}$, and  
      \beq{3.18}
        \vec{\lambda}_{s} \vec{\lambda}_{l} = 
            \frac{1}{2} K_l A_{sl}  - \frac{D - 3}{D -2} \equiv G_{sl},
      \eeq    
     $s,l = 1, 2, 3, 4$;  (\ref{3.17}) is a special case of  (\ref{3.18}). 
       
       From (\ref{2.16}),(\ref{2.18}) it also follows that 
       \beq{3.26}
       \frac{h_s}{h_l} = \frac{K_{l}}{K_{s}} = \frac{B_{ll}}{B_{ss}} = 
          \frac{B_{ls}}{B_{ss}} \frac{B_{ll}}{B_{sl}} = \frac{A_{ls}}{A_{sl}}
        \eeq
       for any $s \neq l$ obeying $A_{sl}, A_{ls} \neq 0$. This
       implies
       \beq{3.26K}
        K_1 =K_2 = K_3 = K, \qquad K_4 = K, \ \frac{1}{2} K, \ 2K, \ K, 
        \eeq
        or 
        \beq{3.26hF}
        h_1 = h_2 = h_3 =  h, \qquad h_4 = h, \ 2h, \ \frac{1}{2} h , \ h   , 
        \eeq
        ($h = K^{-1}$) for ${\cal G} =  A_4,  B_4,  C_4, D_4$,  respectively,
        and 
        \beq{3.26KF}
               K_1 =K_2 =  K, \qquad K_3 = K_4 = \frac{1}{2} K,  
                \eeq
                or 
                \beq{3.26h}
                h_1 = h_2 =  h, \qquad h_3 = h_4 =  2h, 
                                \eeq
        ($h = K^{-1}$) for ${\cal G} =  F_4$.

     
     {\bf Polynomials.}
     According to the polynomial conjecture, the set of moduli functions  $H_1(z), ..., H_4(z)$, obeying eqs. (\ref{1.1}) and (\ref{1.2})
     with the Cartan matrix $A =  (A_{sl})$ from (\ref{A.5}) 
     are polynomials with powers 
     \beq{3.n}
     (n_1,n_2, n_3, n_4) = (4,6,6,4), (8,14,18,10), (7,12,15,16), (6,10,6,6), (22,42,30,16) 
     \eeq 
     calculated by using (\ref{1.4}) for
      Lie algebras $A_4$, $B_4$, $C_4$, $D_4$, $F_4$, respectively.
     
     One can prove this conjecture by solving the system of non-linear algebraic equations for the coefficients of these polynomials following from master equations (\ref{1.1}). Below we present a list of the polynomials obtained by using appropriate MATHEMATICA algorithm. For convenience,
     we use the rescaled variables (as in ref. \cite{I-14}):
     \beq{3.P}
        p_s = P_s/n_s.    
     \eeq
              
          {\bf $A_4$-case.} For the Lie algebra $A_4 \cong sl(5)$  we have
\bearr{A.4}
H_1=1+4 p_1 \textcolor{red}{z}+6 p_1 p_2 \textcolor{red}{z^2}+4 p_1 p_2 p_3 \textcolor{red}{z^3}+p_1 p_2 p_3 p_4 \textcolor{red}{z^4},
\nl
H_2=1+6 p_2 \textcolor{red}{z}+\left(6 p_1 p_2+9 p_2 p_3\right) \textcolor{red}{z^2}+\left(16 p_1 p_2 p_3+4 p_2 p_3 p_4\right) \textcolor{red}{z^3}\nll
+\left(6 p_1 p_2^2 p_3+9 p_1 p_2 p_3 p_4\right) \textcolor{red}{z^4}+6 p_1 p_2^2 p_3 p_4 \textcolor{red}{z^5}+p_1 p_2^2 p_3^2 p_4 \textcolor{red}{z^6},
\nl
H_3=1+6 p_3 \textcolor{red}{z}+\left(9 p_2 p_3+6 p_3 p_4\right) \textcolor{red}{z^2}+\left(4 p_1 p_2 p_3+16 p_2 p_3 p_4\right) \textcolor{red}{z^3}\nll
+\left(9 p_1 p_2 p_3 p_4+6 p_2 p_3^2 p_4\right) \textcolor{red}{z^4}+6 p_1 p_2 p_3^2 p_4 \textcolor{red}{z^5}+p_1 p_2^2 p_3^2 p_4 \textcolor{red}{z^6},
\nl
H_4=1+4 p_4 \textcolor{red}{z}+6 p_3 p_4 \textcolor{red}{z^2}+4 p_2 p_3 p_4 \textcolor{red}{z^3}+p_1 p_2 p_3 p_4 \textcolor{red}{z^4}.
\ear
                     {\bf $B_4$-case.}
For the Lie algebra $B_4 \cong so(9)$  the fluxbrane polynomials are:
\bearr{B.4}
H_1=1+8 p_1 \textcolor{red}{z}+28 p_1 p_2 \textcolor{red}{z^2}+56 p_1 p_2 p_3 \textcolor{red}{z^3}+70 p_1 p_2 p_3 p_4 \textcolor{red}{z^4}+56 p_1 p_2 p_3 p_4^2 \textcolor{red}{z^5}\nll
+28 p_1 p_2 p_3^2 p_4^2 \textcolor{red}{z^6}+8 p_1 p_2^2 p_3^2 p_4^2 \textcolor{red}{z^7}+p_1^2 p_2^2 p_3^2 p_4^2 \textcolor{red}{z^8},
\nl
H_2=1+14 p_2 \textcolor{red}{z}+\left(28 p_1 p_2+63 p_2 p_3\right) \textcolor{red}{z^2}+\left(224 p_1 p_2 p_3+140 p_2 p_3 p_4\right) \textcolor{red}{z^3}\nll
+(196 p_1 p_2^2 p_3+630 p_1 p_2 p_3 p_4+175 p_2 p_3 p_4^2) \textcolor{red}{z^4}\nll
+\left(980 p_1 p_2^2 p_3 p_4+896 p_1 p_2 p_3 p_4^2+126 p_2 p_3^2 p_4^2\right) \textcolor{red}{z^5}\nll
+\left(490 p_1 p_2^2 p_3^2 p_4+1764 p_1 p_2^2 p_3 p_4^2+700 p_1 p_2 p_3^2 p_4^2+49 p_2^2 p_3^2 p_4^2\right) \textcolor{red}{z^6}+3432 p_1 p_2^2 p_3^2 p_4^2 \textcolor{red}{z^7}\nll
+\left(49 p_1^2 p_2^2 p_3^2 p_4^2+700 p_1 p_2^3 p_3^2 p_4^2+1764 p_1 p_2^2 p_3^3 p_4^2+490 p_1 p_2^2 p_3^2 p_4^3\right) \textcolor{red}{z^8}\nll
+\left(126 p_1^2 p_2^3 p_3^2 p_4^2+896 p_1 p_2^3 p_3^3 p_4^2+980 p_1 p_2^2 p_3^3 p_4^3\right) \textcolor{red}{z^9}\nll
+\left(175 p_1^2 p_2^3 p_3^3 p_4^2+630 p_1 p_2^3 p_3^3 p_4^3+196 p_1 p_2^2 p_3^3 p_4^4\right) \textcolor{red}{z^{10}}+\left(140 p_1^2 p_2^3 p_3^3 p_4^3+224 p_1 p_2^3 p_3^3 p_4^4\right) \textcolor{red}{z^{11}}\nll
+\left(63 p_1^2 p_2^3 p_3^3 p_4^4+28 p_1 p_2^3 p_3^4 p_4^4\right) \textcolor{red}{z^{12}}+14 p_1^2 p_2^3 p_3^4 p_4^4 \textcolor{red}{z^{13}}+p_1^2 p_2^4 p_3^4 p_4^4 \textcolor{red}{z^{14}},
\nl
H_3=1+18 p_3 \textcolor{red}{z}+\left(63 p_2 p_3+90 p_3 p_4\right) \textcolor{red}{z^2}+\left(56 p_1 p_2 p_3+560 p_2 p_3 p_4+200 p_3 p_4^2\right) \textcolor{red}{z^3}\nll
+\left(630 p_1 p_2 p_3 p_4+630 p_2 p_3^2 p_4+1575 p_2 p_3 p_4^2+225 p_3^2 p_4^2\right) \textcolor{red}{z^4}\nll
+\left(1260 p_1 p_2 p_3^2 p_4+2016 p_1 p_2 p_3 p_4^2+5292 p_2 p_3^2 p_4^2\right) \textcolor{red}{z^5}\nll
+\left(490 p_1 p_2^2 p_3^2 p_4+9996 p_1 p_2 p_3^2 p_4^2+1225 p_2^2 p_3^2 p_4^2+5103 p_2 p_3^3 p_4^2+1750 p_2 p_3^2 p_4^3\right) \textcolor{red}{z^6}\nll
+\left(5616 p_1 p_2^2 p_3^2 p_4^2+12600 p_1 p_2 p_3^3 p_4^2+3528 p_2^2 p_3^3 p_4^2+5040 p_1 p_2 p_3^2 p_4^3+5040 p_2 p_3^3 p_4^3\right) \textcolor{red}{z^7}\nll
+\left(441 p_1^2 p_2^2 p_3^2 p_4^2+17172 p_1 p_2^2 p_3^3 p_4^2+4410 p_1 p_2^2 p_3^2 p_4^3+15750 p_1 p_2 p_3^3 p_4^3+4410 p_2^2 p_3^3 p_4^3+1575 p_2 p_3^3 p_4^4\right) \textcolor{red}{z^8}\nll
+\left(2450 p_1^2 p_2^2 p_3^3 p_4^2+5600 p_1 p_2^3 p_3^3 p_4^2+32520 p_1 p_2^2 p_3^3 p_4^3+5600 p_1 p_2 p_3^3 p_4^4+2450 p_2^2 p_3^3 p_4^4\right) \textcolor{red}{z^9}\nll
+\left(1575 p_1^2 p_2^3 p_3^3 p_4^2+4410 p_1^2 p_2^2 p_3^3 p_4^3+15750 p_1 p_2^3 p_3^3 p_4^3+4410 p_1 p_2^2 p_3^4 p_4^3+17172 p_1 p_2^2 p_3^3 p_4^4+441 p_2^2 p_3^4 p_4^4\right) \textcolor{red}{z^{10}}\nll
+\left(5040 p_1^2 p_2^3 p_3^3 p_4^3+5040 p_1 p_2^3 p_3^4 p_4^3+3528 p_1^2 p_2^2 p_3^3 p_4^4+12600 p_1 p_2^3 p_3^3 p_4^4+5616 p_1 p_2^2 p_3^4 p_4^4\right) \textcolor{red}{z^{11}}\nll
+\left(1750 p_1^2 p_2^3 p_3^4 p_4^3+5103 p_1^2 p_2^3 p_3^3 p_4^4+1225 p_1^2 p_2^2 p_3^4 p_4^4+9996 p_1 p_2^3 p_3^4 p_4^4+490 p_1 p_2^2 p_3^4 p_4^5\right) \textcolor{red}{z^{12}}\nll
+\left(5292 p_1^2 p_2^3 p_3^4 p_4^4+2016 p_1 p_2^3 p_3^5 p_4^4+1260 p_1 p_2^3 p_3^4 p_4^5\right) \textcolor{red}{z^{13}}\nll
+\left(225 p_1^2 p_2^4 p_3^4 p_4^4+1575 p_1^2 p_2^3 p_3^5 p_4^4+630 p_1^2 p_2^3 p_3^4 p_4^5+630 p_1 p_2^3 p_3^5 p_4^5\right) \textcolor{red}{z^{14}}\nll
+\left(200 p_1^2 p_2^4 p_3^5 p_4^4+560 p_1^2 p_2^3 p_3^5 p_4^5+56 p_1 p_2^3 p_3^5 p_4^6\right) \textcolor{red}{z^{15}}\nll
+\left(90 p_1^2 p_2^4 p_3^5 p_4^5+63 p_1^2 p_2^3 p_3^5 p_4^6\right) \textcolor{red}{z^{16}}+18 p_1^2 p_2^4 p_3^5 p_4^6 \textcolor{red}{z^{17}}+p_1^2 p_2^4 p_3^6 p_4^6 \textcolor{red}{z^{18}},
\nl
H_4=1+10 p_4 \textcolor{red}{z}+45 p_3 p_4 \textcolor{red}{z^2}+\left(70 p_2 p_3 p_4+50 p_3 p_4^2\right) \textcolor{red}{z^3}+\left(35 p_1 p_2 p_3 p_4+175 p_2 p_3 p_4^2\right) \textcolor{red}{z^4}\nll
+\left(126 p_1 p_2 p_3 p_4^2+126 p_2 p_3^2 p_4^2\right) \textcolor{red}{z^5}+\left(175 p_1 p_2 p_3^2 p_4^2+35 p_2 p_3^2 p_4^3\right) \textcolor{red}{z^6}\nll
+\left(50 p_1 p_2^2 p_3^2 p_4^2+70 p_1 p_2 p_3^2 p_4^3\right) \textcolor{red}{z^7}+45 p_1 p_2^2 p_3^2 p_4^3 \textcolor{red}{z^8}+10 p_1 p_2^2 p_3^3 p_4^3 \textcolor{red}{z^9}+p_1 p_2^2 p_3^3 p_4^4 \textcolor{red}{z^{10}}.
\ear
   
 \newpage                         
{ \bf $C_4$-case. }
           For the Lie algebra $C_4 \cong sp(6)$ we get the following polynomials
           \vspace{10pt}
\bearr{C.4}
H_1=1+7 p_1 \textcolor{red}{z}+21 p_1 p_2 \textcolor{red}{z^2}+35 p_1 p_2 p_3 \textcolor{red}{z^3}+35 p_1 p_2 p_3 p_4 \textcolor{red}{z^4}\nll
+21 p_1 p_2 p_3^2 p_4 \textcolor{red}{z^5}+7 p_1 p_2^2 p_3^2 p_4 \textcolor{red}{z^6}+p_1^2 p_2^2 p_3^2 p_4 \textcolor{red}{z^7},
\nl
H_2=1+12 p_2 \textcolor{red}{z}+\left(21 p_1 p_2+45 p_2 p_3\right) \textcolor{red}{z^2}+\left(140 p_1 p_2 p_3+80 p_2 p_3 p_4\right) \textcolor{red}{z^3}\nll
+\left(105 p_1 p_2^2 p_3+315 p_1 p_2 p_3 p_4+75 p_2 p_3^2 p_4\right) \textcolor{red}{z^4}+\left(420 p_1 p_2^2 p_3 p_4+336 p_1 p_2 p_3^2 p_4+36 p_2^2 p_3^2 p_4\right) \textcolor{red}{z^5}\nll
+924 p_1 p_2^2 p_3^2 p_4 \textcolor{red}{z^6}+\left(36 p_1^2 p_2^2 p_3^2 p_4+336 p_1 p_2^3 p_3^2 p_4+420 p_1 p_2^2 p_3^3 p_4\right) \textcolor{red}{z^7}\nll
+\left(75 p_1^2 p_2^3 p_3^2 p_4+315 p_1 p_2^3 p_3^3 p_4+105 p_1 p_2^2 p_3^3 p_4^2\right) \textcolor{red}{z^8}+\left(80 p_1^2 p_2^3 p_3^3 p_4+140 p_1 p_2^3 p_3^3 p_4^2\right) \textcolor{red}{z^9}\nll
+\left(45 p_1^2 p_2^3 p_3^3 p_4^2+21 p_1 p_2^3 p_3^4 p_4^2\right) \textcolor{red}{z^{10}}+12 p_1^2 p_2^3 p_3^4 p_4^2 \textcolor{red}{z^{11}}+p_1^2 p_2^4 p_3^4 p_4^2 \textcolor{red}{z^{12}},
\nl
H_3=1+15 p_3 \textcolor{red}{z}+\left(45 p_2 p_3+60 p_3 p_4\right) \textcolor{red}{z^2}+\left(35 p_1 p_2 p_3+320 p_2 p_3 p_4+100 p_3^2 p_4\right) \textcolor{red}{z^3}\nll
+\left(315 p_1 p_2 p_3 p_4+1050 p_2 p_3^2 p_4\right) \textcolor{red}{z^4}+\left(1302 p_1 p_2 p_3^2 p_4+576 p_2^2 p_3^2 p_4+1125 p_2 p_3^3 p_4\right) \textcolor{red}{z^5}\nll
+\left(1050 p_1 p_2^2 p_3^2 p_4+2240 p_1 p_2 p_3^3 p_4+1215 p_2^2 p_3^3 p_4+500 p_2 p_3^3 p_4^2\right)\textcolor{red}{z^6}
\nll
+\left(225 p_1^2 p_2^2 p_3^2 p_4+3990 p_1 p_2^2 p_3^3 p_4+1260 p_1 p_2 p_3^3 p_4^2+960 p_2^2 p_3^3 p_4^2\right) \textcolor{red}{z^7}\nll
+\left(960 p_1^2 p_2^2 p_3^3 p_4+1260 p_1 p_2^3 p_3^3 p_4+3990 p_1 p_2^2 p_3^3 p_4^2+225 p_2^2 p_3^4 p_4^2\right) \textcolor{red}{z^8}\nll
+\left(500 p_1^2 p_2^3 p_3^3 p_4+1215 p_1^2 p_2^2 p_3^3 p_4^2+2240 p_1 p_2^3 p_3^3 p_4^2+1050 p_1 p_2^2 p_3^4 p_4^2\right) \textcolor{red}{z^9}\nll
+\left(1125 p_1^2 p_2^3 p_3^3 p_4^2+576 p_1^2 p_2^2 p_3^4 p_4^2+1302 p_1 p_2^3 p_3^4 p_4^2\right) \textcolor{red}{z^{10}}+\left(1050 p_1^2 p_2^3 p_3^4 p_4^2+315 p_1 p_2^3 p_3^5 p_4^2\right) \textcolor{red}{z^{11}}\nll
+\left(100 p_1^2 p_2^4 p_3^4 p_4^2+320 p_1^2 p_2^3 p_3^5 p_4^2+35 p_1 p_2^3 p_3^5 p_4^3\right) \textcolor{red}{z^{12}}+\left(60 p_1^2 p_2^4 p_3^5 p_4^2+45 p_1^2 p_2^3 p_3^5 p_4^3\right) \textcolor{red}{z^{13}}\nll
+15 p_1^2 p_2^4 p_3^5 p_4^3 \textcolor{red}{z^{14}}+p_1^2 p_2^4 p_3^6 p_4^3 \textcolor{red}{z^{15}},
\nl
H_4=1+16 p_4 \textcolor{red}{z}+120 p_3 p_4 \textcolor{red}{z^2}+\left(160 p_2 p_3 p_4+400 p_3^2 p_4\right) \textcolor{red}{z^3}\nll
+\left(70 p_1 p_2 p_3 p_4+1350 p_2 p_3^2 p_4+400 p_3^2 p_4^2\right) \textcolor{red}{z^4}+\left(672 p_1 p_2 p_3^2 p_4+1296 p_2^2 p_3^2 p_4+2400 p_2 p_3^2 p_4^2\right) \textcolor{red}{z^5}\nll
+\left(1400 p_1 p_2^2 p_3^2 p_4+1512 p_1 p_2 p_3^2 p_4^2+4096 p_2^2 p_3^2 p_4^2+1000 p_2 p_3^3 p_4^2\right) \textcolor{red}{z^6}\nll
+\left(400 p_1^2 p_2^2 p_3^2 p_4+5600 p_1 p_2^2 p_3^2 p_4^2+1120 p_1 p_2 p_3^3 p_4^2+4320 p_2^2 p_3^3 p_4^2\right) \textcolor{red}{z^7}\nll
+\left(2025 p_1^2 p_2^2 p_3^2 p_4^2+8820 p_1 p_2^2 p_3^3 p_4^2+2025 p_2^2 p_3^4 p_4^2\right) \textcolor{red}{z^8}\nll
+\left(4320 p_1^2 p_2^2 p_3^3 p_4^2+1120 p_1 p_2^3 p_3^3 p_4^2+5600 p_1 p_2^2 p_3^4 p_4^2+400 p_2^2 p_3^4 p_4^3\right) \textcolor{red}{z^9}\nll
+\left(1000 p_1^2 p_2^3 p_3^3 p_4^2+4096 p_1^2 p_2^2 p_3^4 p_4^2+1512 p_1 p_2^3 p_3^4 p_4^2+1400 p_1 p_2^2 p_3^4 p_4^3\right) \textcolor{red}{z^{10}}\nll
+\left(2400 p_1^2 p_2^3 p_3^4 p_4^2+1296 p_1^2 p_2^2 p_3^4 p_4^3+672 p_1 p_2^3 p_3^4 p_4^3\right) \textcolor{red}{z^{11}}\nll
+\left(400 p_1^2 p_2^4 p_3^4 p_4^2+1350 p_1^2 p_2^3 p_3^4 p_4^3+70 p_1 p_2^3 p_3^5 p_4^3\right) \textcolor{red}{z^{12}}+\left(400 p_1^2 p_2^4 p_3^4 p_4^3+160 p_1^2 p_2^3 p_3^5 p_4^3\right) \textcolor{red}{z^{13}}\nll
+120 p_1^2 p_2^4 p_3^5 p_4^3 \textcolor{red}{z^{14}}+16 p_1^2 p_2^4 p_3^6 p_4^3 \textcolor{red}{z^{15}}+p_1^2 p_2^4 p_3^6 p_4^4 \textcolor{red}{z^{16}}.
\ear
  
   { \bf $D_4$-case. }
            For the Lie algebra $D_4 \cong so(8)$ we obtain the polynomials
            \vspace{10pt}
 \bearr{D.4}
H_1=1+6 p_1 \textcolor{red}{z}+15 p_1 p_2 \textcolor{red}{z^2}+\left(10 p_1 p_2 p_3+10 p_1 p_2 p_4\right) \textcolor{red}{z^3}+15 p_1 p_2 p_3 p_4 \textcolor{red}{z^4}\nll
+6 p_1 p_2^2 p_3 p_4 \textcolor{red}{z^5}+p_1^2 p_2^2 p_3 p_4 \textcolor{red}{z^6},
\nl
H_2=1+10 p_2 \textcolor{red}{z}+\left(15 p_1 p_2+15 p_2 p_3+15 p_2 p_4\right) \textcolor{red}{z^2}+\left(40 p_1 p_2 p_3+40 p_1 p_2 p_4+40 p_2 p_3 p_4\right) \textcolor{red}{z^3}\nll
+\left(25 p_1 p_2^2 p_3+25 p_1 p_2^2 p_4+135 p_1 p_2 p_3 p_4+25 p_2^2 p_3 p_4\right) \textcolor{red}{z^4}+252 p_1 p_2^2 p_3 p_4 \textcolor{red}{z^5}\nll
+\left(25 p_1^2 p_2^2 p_3 p_4+135 p_1 p_2^3 p_3 p_4+25 p_1 p_2^2 p_3^2 p_4+25 p_1 p_2^2 p_3 p_4^2\right) \textcolor{red}{z^6}\nll
+\left(40 p_1^2 p_2^3 p_3 p_4+40 p_1 p_2^3 p_3^2 p_4+40 p_1 p_2^3 p_3 p_4^2\right) \textcolor{red}{z^7}+\left(15 p_1^2 p_2^3 p_3^2 p_4+15 p_1^2 p_2^3 p_3 p_4^2+15 p_1 p_2^3 p_3^2 p_4^2\right) \textcolor{red}{z^8}\nll+10 p_1^2 p_2^3 p_3^2 p_4^2 \textcolor{red}{z^9}+p_1^2 p_2^4 p_3^2 p_4^2 \textcolor{red}{z^{10}},
\nl
H_3=1+6 p_3 \textcolor{red}{z}+15 p_2 p_3 \textcolor{red}{z^2}+\left(10 p_1 p_2 p_3+10 p_2 p_3 p_4\right) \textcolor{red}{z^3}+15 p_1 p_2 p_3 p_4 \textcolor{red}{z^4}\nll
+6 p_1 p_2^2 p_3 p_4 \textcolor{red}{z^5}+p_1 p_2^2 p_3^2 p_4 \textcolor{red}{z^6},
\nl
H_4=1+6 p_4 \textcolor{red}{z}+15 p_2 p_4 \textcolor{red}{z^2}+\left(10 p_1 p_2 p_4+10 p_2 p_3 p_4\right) \textcolor{red}{z^3}+15 p_1 p_2 p_3 p_4 \textcolor{red}{z^4}\nll
+6 p_1 p_2^2 p_3 p_4 \textcolor{red}{z^5}+p_1 p_2^2 p_3 p_4^2 \textcolor{red}{z^6}.
\ear 
 
 \newpage
 
 { \bf $F_4$-case.}      
     For the exceptional Lie algebra $F_4$ we find the following polynomials: 
              \vspace{15pt}         

\begin{itemize}
\item[$H_1=$]\(
1+22 p_1 \textcolor{red}{z}+231 p_1 p_2 \textcolor{red}{z^2}+1540 p_1 p_2 p_3 \textcolor{red}{z^3}+(5775 p_1 p_2 p_3^2+1540 p_1 p_2 p_3 p_4) \textcolor{red}{z^4}+(9702 p_1 p_2^2 p_3^2+16632 p_1 p_2 p_3^2 p_4) \textcolor{red}{z^5} 
+(5929 p_1^2 p_2^2 p_3^2+53900 p_1 p_2^2 p_3^2 p_4+14784 p_1 p_2 p_3^2 p_4^2) \textcolor{red}{z^6}+(47432 p_1^2 p_2^2 p_3^2 p_4+33000 p_1 p_2^2 p_3^3 p_4+90112 p_1 p_2^2 p_3^2 p_4^2) \textcolor{red}{z^7} 
+(65340 p_1^2 p_2^2 p_3^3 p_4+108900 p_1^2 p_2^2 p_3^2 p_4^2+145530 p_1 p_2^2 p_3^3 p_4^2) \textcolor{red}{z^8}+(33880 p_1^2 p_2^3 p_3^3 p_4+355740 p_1^2 p_2^2 p_3^3 p_4^2+107800 p_1 p_2^2 p_3^4 p_4^2) \textcolor{red}{z^9} 
+(10164 p_1^2 p_2^3 p_3^4 p_4+211750 p_1^2 p_2^3 p_3^3 p_4^2+379456 p_1^2 p_2^2 p_3^4 p_4^2+45276 p_1 p_2^3 p_3^4 p_4^2) \textcolor{red}{z^{10}}+705432 p_1^2 p_2^3 p_3^4 p_4^2 \textcolor{red}{z^{11}} 
+(45276 p_1^3 p_2^3 p_3^4 p_4^2+379456 p_1^2 p_2^4 p_3^4 p_4^2+211750 p_1^2 p_2^3 p_3^5 p_4^2+10164 p_1^2 p_2^3 p_3^4 p_4^3) \textcolor{red}{z^{12}} 
+(107800 p_1^3 p_2^4 p_3^4 p_4^2+355740 p_1^2 p_2^4 p_3^5 p_4^2+33880 p_1^2 p_2^3 p_3^5 p_4^3) \textcolor{red}{z^{13}} 
+(145530 p_1^3 p_2^4 p_3^5 p_4^2+108900 p_1^2 p_2^4 p_3^6 p_4^2+65340 p_1^2 p_2^4 p_3^5 p_4^3) \textcolor{red}{z^{14}} 
+(90112 p_1^3 p_2^4 p_3^6 p_4^2+33000 p_1^3 p_2^4 p_3^5 p_4^3+47432 p_1^2 p_2^4 p_3^6 p_4^3) \textcolor{red}{z^{15}}+(14784 p_1^3 p_2^5 p_3^6 p_4^2+53900 p_1^3 p_2^4 p_3^6 p_4^3+5929 p_1^2 p_2^4 p_3^6 p_4^4) \textcolor{red}{z^{16}} 
+(16632 p_1^3 p_2^5 p_3^6 p_4^3+9702 p_1^3 p_2^4 p_3^6 p_4^4) \textcolor{red}{z^{17}}+(1540 p_1^3 p_2^5 p_3^7 p_4^3+5775 p_1^3 p_2^5 p_3^6 p_4^4) \textcolor{red}{z^{18}}+1540 p_1^3 p_2^5 p_3^7 p_4^4 \textcolor{red}{z^{19}}+231 p_1^3 p_2^5 p_3^8 p_4^4 \textcolor{red}{z^{20}} 
+22 p_1^3 p_2^6 p_3^8 p_4^4 \textcolor{red}{z^{21}}+p_1^4 p_2^6 p_3^8 p_4^4 \textcolor{red}{z^{22}},
\)
\begin{equation}\label{H1}\end{equation}
\item[$H_2=$]\(
1+42 p_2 \textcolor{red}{z}+(231 p_1 p_2+630 p_2 p_3) \textcolor{red}{z^2}+(6160 p_1 p_2 p_3+4200 p_2 p_3^2+1120 p_2 p_3 p_4) \textcolor{red}{z^3}   +(16170 p_1 p_2^2 p_3+51975 p_1 p_2 p_3^2+11025 p_2^2 p_3^2+13860 p_1 p_2 p_3 p_4+18900 p_2 p_3^2 p_4) \textcolor{red}{z^4}
+(407484 p_1 p_2^2 p_3^2+64680 p_1 p_2^2 p_3 p_4+266112 p_1 p_2 p_3^2 p_4+88200 p_2^2 p_3^2 p_4+24192 p_2 p_3^2 p_4^2) \textcolor{red}{z^5}
+(148225 p_1^2 p_2^2 p_3^2+916839 p_1 p_2^3 p_3^2+404250 p_1 p_2^2 p_3^3+3132668 p_1 p_2^2 p_3^2 p_4 +73500 p_2^2 p_3^3 p_4+369600 p_1 p_2 p_3^2 p_4^2+200704 p_2^2 p_3^2 p_4^2) \textcolor{red}{z^6} 
+(996072 p_1^2 p_2^3 p_3^2+2716560 p_1 p_2^3 p_3^3+1707552 p_1^2 p_2^2 p_3^2 p_4+9055200 p_1 p_2^3 p_3^2 p_4 +6035040 p_1 p_2^2 p_3^3 p_4+6044544 p_1 p_2^2 p_3^2 p_4^2+423360 p_2^2 p_3^3 p_4^2) \textcolor{red}{z^7}
+(3735270 p_1^2 p_2^3 p_3^3+2546775 p_1 p_2^3 p_3^4+12450900 p_1^2 p_2^3 p_3^2 p_4+3201660 p_1^2 p_2^2 p_3^3 p_4+43423380 p_1 p_2^3 p_3^3 p_4+4365900 p_1 p_2^2 p_3^4 p_4 +5336100 p_1^2 p_2^2 p_3^2 p_4^2+23654400 p_1 p_2^3 p_3^2 p_4^2+18918900 p_1 p_2^2 p_3^3 p_4^2+396900 p_2^2 p_3^4 p_4^2) \textcolor{red}{z^8} 
+(6225450 p_1^2 p_2^3 p_3^4+81650800 p_1^2 p_2^3 p_3^3 p_4+93601200 p_1 p_2^3 p_3^4 p_4+41164200 p_1^2 p_2^3 p_3^2 p_4^2+22767360 p_1^2 p_2^2 p_3^3 p_4^2+171990280 p_1 p_2^3 p_3^3 p_4^2+24147200 p_1 p_2^2 p_3^4 p_4^2+205800 p_2^3 p_3^4 p_4^2+4139520 p_1 p_2^2 p_3^3 p_4^3) \textcolor{red}{z^9}
+(2614689 p_1^2 p_2^4 p_3^4+17431260 p_1^2 p_2^4 p_3^3 p_4+231708708 p_1^2 p_2^3 p_3^4 p_4+23769900 p_1 p_2^4 p_3^4 p_4+77962500 p_1 p_2^3 p_3^5 p_4 +420637140 p_1^2 p_2^3 p_3^3 p_4^2+30735936 p_1^2 p_2^2 p_3^4 p_4^2+598635576 p_1 p_2^3 p_3^4 p_4^2+56770560 p_1 p_2^3 p_3^3 p_4^3+11176704 p_1 p_2^2 p_3^4 p_4^3) \textcolor{red}{z^{10}}
+(175877856 p_1^2 p_2^4 p_3^4 p_4+274428000 p_1^2 p_2^3 p_3^5 p_4+58212000 p_1 p_2^4 p_3^5 p_4+142296000 p_1^2 p_2^4 p_3^3 p_4^2+1896293952 p_1^2 p_2^3 p_3^4 p_4^2 +191866752 p_1 p_2^4 p_3^4 p_4^2+984060000 p_1 p_2^3 p_3^5 p_4^2+121968000 p_1^2 p_2^3 p_3^3 p_4^3+435558816 p_1 p_2^3 p_3^4 p_4^3) \textcolor{red}{z^{11}}
+(12782924 p_1^3 p_2^4 p_3^4 p_4+525427980 p_1^2 p_2^4 p_3^5 p_4+5478396 p_1^3 p_2^3 p_3^4 p_4^2+2005022376 p_1^2 p_2^4 p_3^4 p_4^2 +4106272940 p_1^2 p_2^3 p_3^5 p_4^2+816487980 p_1 p_2^4 p_3^5 p_4^2+707437500 p_1 p_2^3 p_3^6 p_4^2+1396604748 p_1^2 p_2^3 p_3^4 p_4^3 +220774400 p_1 p_2^4 p_3^4 p_4^3+1201272380 p_1 p_2^3 p_3^5 p_4^3+60555264 p_1 p_2^3 p_3^4 p_4^4) \textcolor{red}{z^{12}}
+(70436520 p_1^3 p_2^4 p_3^5 p_4+239057280 p_1^2 p_2^5 p_3^5 p_4+96049800 p_1^2 p_2^4 p_3^6 p_4+180457200 p_1^3 p_2^4 p_3^4 p_4^2 +398428800 p_1^2 p_2^5 p_3^4 p_4^2+9178974000 p_1^2 p_2^4 p_3^5 p_4^2+3585859200 p_1^2 p_2^3 p_3^6 p_4^2+1189465200 p_1 p_2^4 p_3^6 p_4^2 +1611502200 p_1^2 p_2^4 p_3^4 p_4^3+5439772800 p_1^2 p_2^3 p_3^5 p_4^3+1540871640 p_1 p_2^4 p_3^5 p_4^3+1303948800 p_1 p_2^3 p_3^6 p_4^3 +292723200 p_1^2 p_2^3 p_3^4 p_4^4+391184640 p_1 p_2^3 p_3^5 p_4^4) \textcolor{red}{z^{13}}
+(82175940 p_1^3 p_2^5 p_3^5 p_4+112058100 p_1^2 p_2^5 p_3^6 p_4+136959900 p_1^3 p_2^5 p_3^4 p_4^2+1285029900 p_1^3 p_2^4 p_3^5 p_4^2 +5685080940 p_1^2 p_2^5 p_3^5 p_4^2+15028648200 p_1^2 p_2^4 p_3^6 p_4^2+499167900 p_1 p_2^5 p_3^6 p_4^2+234788400 p_1^3 p_2^4 p_3^4 p_4^3 +15327479700 p_1^2 p_2^4 p_3^5 p_4^3+7171718400 p_1^2 p_2^3 p_3^6 p_4^3+3451486500 p_1 p_2^4 p_3^6 p_4^3+446054400 p_1^2 p_2^4 p_3^4 p_4^4+2151515520 p_1^2 p_2^3 p_3^5 p_4^4+596090880 p_1 p_2^4 p_3^5 p_4^4+651974400 p_1 p_2^3 p_3^6 p_4^4) \textcolor{red}{z^{14}} 
+(43827168 p_1^3 p_2^5 p_3^6 p_4+2179888480 p_1^3 p_2^5 p_3^5 p_4^2+2414513024 p_1^3 p_2^4 p_3^6 p_4^2+21026246976 p_1^2 p_2^5 p_3^6 p_4^2 +3557400000 p_1^2 p_2^4 p_3^7 p_4^2+3277206240 p_1^3 p_2^4 p_3^5 p_4^3+10654446880 p_1^2 p_2^5 p_3^5 p_4^3 +38613582112 p_1^2 p_2^4 p_3^6 p_4^3+1774819200 p_1 p_2^5 p_3^6 p_4^3+646800000 p_1 p_2^4 p_3^7 p_4^3+8150714880 p_1^2 p_2^4 p_3^5 p_4^4 +4079910912 p_1^2 p_2^3 p_3^6 p_4^4+2253071744 p_1 p_2^4 p_3^6 p_4^4) \textcolor{red}{z^{15}}
+(9717029784 p_1^3 p_2^5 p_3^6 p_4^2+8199664704 p_1^2 p_2^6 p_3^6 p_4^2+13199224500 p_1^2 p_2^5 p_3^7 p_4^2+4946287500 p_1^3 p_2^5 p_3^5 p_4^3+10108843668 p_1^3 p_2^4 p_3^6 p_4^3+64474736508 p_1^2 p_2^5 p_3^6 p_4^3+14007262500 p_1^2 p_2^4 p_3^7 p_4^3+611226000 p_1 p_2^5 p_3^7 p_4^3+1760913000 p_1^3 p_2^4 p_3^5 p_4^4+7805952000 p_1^2 p_2^5 p_3^5 p_4^4+29296429974 p_1^2 p_2^4 p_3^6 p_4^4+1669054464 p_1 p_2^5 p_3^6 p_4^4+713097000 p_1 p_2^4 p_3^7 p_4^4) \textcolor{red}{z^{16}}
+(439267752 p_1^4 p_2^5 p_3^6 p_4^2+6754454784 p_1^3 p_2^6 p_3^6 p_4^2+6903638280 p_1^3 p_2^5 p_3^7 p_4^2+10040405760 p_1^2 p_2^6 p_3^7 p_4^2+2858625000 p_1^2 p_2^5 p_3^8 p_4^2+37825702992 p_1^3 p_2^5 p_3^6 p_4^3+33468019200 p_1^2 p_2^6 p_3^6 p_4^3+4507937280 p_1^3 p_2^4 p_3^7 p_4^3+57537501840 p_1^2 p_2^5 p_3^7 p_4^3+4192650000 p_1^3 p_2^5 p_3^5 p_4^4+8611029504 p_1^3 p_2^4 p_3^6 p_4^4+63276492636 p_1^2 p_2^5 p_3^6 p_4^4+16802311680 p_1^2 p_2^4 p_3^7 p_4^4+1198002960 p_1 p_2^5 p_3^7 p_4^4+245887488 p_1^2 p_2^4 p_3^6 p_4^5) \textcolor{red}{z^{17}} 
+(1423552900 p_1^4 p_2^6 p_3^6 p_4^2+10086748980 p_1^3 p_2^6 p_3^7 p_4^2+2862182400 p_1^3 p_2^5 p_3^8 p_4^2+3890016900 p_1^2 p_2^6 p_3^8 p_4^2+2440376400 p_1^4 p_2^5 p_3^6 p_4^3+33759456500 p_1^3 p_2^6 p_3^6 p_4^3+44524657100 p_1^3 p_2^5 p_3^7 p_4^3+59339922180 p_1^2 p_2^6 p_3^7 p_4^3+16165587900 p_1^2 p_2^5 p_3^8 p_4^3+43888833450 p_1^3 p_2^5 p_3^6 p_4^4+38856294400 p_1^2 p_2^6 p_3^6 p_4^4+6135803520 p_1^3 p_2^4 p_3^7 p_4^4+86086107380 p_1^2 p_2^5 p_3^7 p_4^4+1859334400 p_1^2 p_2^4 p_3^8 p_4^4+221852400 p_1 p_2^5 p_3^8 p_4^4+1040793600 p_1^2 p_2^5 p_3^6 p_4^5+1115600640 p_1^2 p_2^4 p_3^7 p_4^5) \textcolor{red}{z^{18}} 
+(2510101440 p_1^4 p_2^6 p_3^7 p_4^2+6411081600 p_1^3 p_2^6 p_3^8 p_4^2+8367004800 p_1^4 p_2^6 p_3^6 p_4^3+2151515520 p_1^4 p_2^5 p_3^7 p_4^3+81592267680 p_1^3 p_2^6 p_3^7 p_4^3+18912247200 p_1^3 p_2^5 p_3^8 p_4^3+38377231200 p_1^2 p_2^6 p_3^8 p_4^3+3585859200 p_1^4 p_2^5 p_3^6 p_4^4+45964195200 p_1^3 p_2^6 p_3^6 p_4^4+79733253600 p_1^3 p_2^5 p_3^7 p_4^4+102862932480 p_1^2 p_2^6 p_3^7 p_4^4+46561158000 p_1^2 p_2^5 p_3^8 p_4^4+804988800 p_1^3 p_2^5 p_3^6 p_4^5+8941474080 p_1^2 p_2^5 p_3^7 p_4^5) \textcolor{red}{z^{19}}
+(1967099904 p_1^4 p_2^6 p_3^8 p_4^2+788889024 p_1^3 p_2^7 p_3^8 p_4^2+24726420180 p_1^4 p_2^6 p_3^7 p_4^3+5259260160 p_1^3 p_2^7 p_3^7 p_4^3+75784320612 p_1^3 p_2^6 p_3^8 p_4^3+8004150000 p_1^2 p_2^6 p_3^9 p_4^3+13340250000 p_1^4 p_2^6 p_3^6 p_4^4+6589016280 p_1^4 p_2^5 p_3^7 p_4^4+166955605740 p_1^3 p_2^6 p_3^7 p_4^4+57761551386 p_1^3 p_2^5 p_3^8 p_4^4+113404704966 p_1^2 p_2^6 p_3^8 p_4^4+9338175000 p_1^2 p_2^5 p_3^9 p_4^4+7582847580 p_1^3 p_2^5 p_3^7 p_4^5+13113999360 p_1^2 p_2^6 p_3^7 p_4^5+9175317228 p_1^2 p_2^5 p_3^8 p_4^5) \textcolor{red}{z^{20}}
+(398428800 p_1^4 p_2^7 p_3^8 p_4^2+2656192000 p_1^4 p_2^7 p_3^7 p_4^3+29530356856 p_1^4 p_2^6 p_3^8 p_4^3+14144946816 p_1^3 p_2^7 p_3^8 p_4^3+20764887000 p_1^3 p_2^6 p_3^9 p_4^3+60120060000 p_1^4 p_2^6 p_3^7 p_4^4+14609056000 p_1^3 p_2^7 p_3^7 p_4^4+3123681792 p_1^4 p_2^5 p_3^8 p_4^4+247562655912 p_1^3 p_2^6 p_3^8 p_4^4+3123681792 p_1^2 p_2^7 p_3^8 p_4^4+14609056000 p_1^3 p_2^5 p_3^9 p_4^4+60120060000 p_1^2 p_2^6 p_3^9 p_4^4+20764887000 p_1^3 p_2^6 p_3^7 p_4^5+14144946816 p_1^3 p_2^5 p_3^8 p_4^5+29530356856 p_1^2 p_2^6 p_3^8 p_4^5+2656192000 p_1^2 p_2^5 p_3^9 p_4^5+398428800 p_1^2 p_2^5 p_3^8 p_4^6) \textcolor{red}{z^{21}}
+(9175317228 p_1^4 p_2^7 p_3^8 p_4^3+13113999360 p_1^4 p_2^6 p_3^9 p_4^3+7582847580 p_1^3 p_2^7 p_3^9 p_4^3+9338175000 p_1^4 p_2^7 p_3^7 p_4^4+113404704966 p_1^4 p_2^6 p_3^8 p_4^4+57761551386 p_1^3 p_2^7 p_3^8 p_4^4+166955605740 p_1^3 p_2^6 p_3^9 p_4^4+6589016280 p_1^2 p_2^7 p_3^9 p_4^4+13340250000 p_1^2 p_2^6 p_3^{10} p_4^4+8004150000 p_1^4 p_2^6 p_3^7 p_4^5+75784320612 p_1^3 p_2^6 p_3^8 p_4^5+5259260160 p_1^3 p_2^5 p_3^9 p_4^5+24726420180 p_1^2 p_2^6 p_3^9 p_4^5+788889024 p_1^3 p_2^5 p_3^8 p_4^6+1967099904 p_1^2 p_2^6 p_3^8 p_4^6) \textcolor{red}{z^{22}}
+(8941474080 p_1^4 p_2^7 p_3^9 p_4^3+804988800 p_1^3 p_2^7 p_3^{10} p_4^3+46561158000 p_1^4 p_2^7 p_3^8 p_4^4+102862932480 p_1^4 p_2^6 p_3^9 p_4^4+79733253600 p_1^3 p_2^7 p_3^9 p_4^4+45964195200 p_1^3 p_2^6 p_3^{10} p_4^4+3585859200 p_1^2 p_2^7 p_3^{10} p_4^4+38377231200 p_1^4 p_2^6 p_3^8 p_4^5+18912247200 p_1^3 p_2^7 p_3^8 p_4^5+81592267680 p_1^3 p_2^6 p_3^9 p_4^5+2151515520 p_1^2 p_2^7 p_3^9 p_4^5+8367004800 p_1^2 p_2^6 p_3^{10} p_4^5+6411081600 p_1^3 p_2^6 p_3^8 p_4^6+2510101440 p_1^2 p_2^6 p_3^9 p_4^6) \textcolor{red}{z^{23}}
+(1115600640 p_1^4 p_2^8 p_3^9 p_4^3+1040793600 p_1^4 p_2^7 p_3^{10} p_4^3+221852400 p_1^5 p_2^7 p_3^8 p_4^4+1859334400 p_1^4 p_2^8 p_3^8 p_4^4+86086107380 p_1^4 p_2^7 p_3^9 p_4^4+6135803520 p_1^3 p_2^8 p_3^9 p_4^4+38856294400 p_1^4 p_2^6 p_3^{10} p_4^4+43888833450 p_1^3 p_2^7 p_3^{10} p_4^4+16165587900 p_1^4 p_2^7 p_3^8 p_4^5+59339922180 p_1^4 p_2^6 p_3^9 p_4^5+44524657100 p_1^3 p_2^7 p_3^9 p_4^5+33759456500 p_1^3 p_2^6 p_3^{10} p_4^5+2440376400 p_1^2 p_2^7 p_3^{10} p_4^5+3890016900 p_1^4 p_2^6 p_3^8 p_4^6+2862182400 p_1^3 p_2^7 p_3^8 p_4^6+10086748980 p_1^3 p_2^6 p_3^9 p_4^6+1423552900 p_1^2 p_2^6 p_3^{10} p_4^6) \textcolor{red}{z^{24}}
+(245887488 p_1^4 p_2^8 p_3^{10} p_4^3+1198002960 p_1^5 p_2^7 p_3^9 p_4^4+16802311680 p_1^4 p_2^8 p_3^9 p_4^4+63276492636 p_1^4 p_2^7 p_3^{10} p_4^4+8611029504 p_1^3 p_2^8 p_3^{10} p_4^4+4192650000 p_1^3 p_2^7 p_3^{11} p_4^4+57537501840 p_1^4 p_2^7 p_3^9 p_4^5+4507937280 p_1^3 p_2^8 p_3^9 p_4^5+33468019200 p_1^4 p_2^6 p_3^{10} p_4^5+37825702992 p_1^3 p_2^7 p_3^{10} p_4^5+2858625000 p_1^4 p_2^7 p_3^8 p_4^6+10040405760 p_1^4 p_2^6 p_3^9 p_4^6+6903638280 p_1^3 p_2^7 p_3^9 p_4^6+6754454784 p_1^3 p_2^6 p_3^{10} p_4^6+439267752 p_1^2 p_2^7 p_3^{10} p_4^6) \textcolor{red}{z^{25}}
+(713097000 p_1^5 p_2^8 p_3^9 p_4^4+1669054464 p_1^5 p_2^7 p_3^{10} p_4^4+29296429974 p_1^4 p_2^8 p_3^{10} p_4^4+7805952000 p_1^4 p_2^7 p_3^{11} p_4^4+1760913000 p_1^3 p_2^8 p_3^{11} p_4^4+611226000 p_1^5 p_2^7 p_3^9 p_4^5+14007262500 p_1^4 p_2^8 p_3^9 p_4^5+64474736508 p_1^4 p_2^7 p_3^{10} p_4^5+10108843668 p_1^3 p_2^8 p_3^{10} p_4^5+4946287500 p_1^3 p_2^7 p_3^{11} p_4^5+13199224500 p_1^4 p_2^7 p_3^9 p_4^6+8199664704 p_1^4 p_2^6 p_3^{10} p_4^6+9717029784 p_1^3 p_2^7 p_3^{10} p_4^6) \textcolor{red}{z^{26}} 
+(2253071744 p_1^5 p_2^8 p_3^{10} p_4^4+4079910912 p_1^4 p_2^9 p_3^{10} p_4^4+8150714880 p_1^4 p_2^8 p_3^{11} p_4^4+646800000 p_1^5 p_2^8 p_3^9 p_4^5+1774819200 p_1^5 p_2^7 p_3^{10} p_4^5+38613582112 p_1^4 p_2^8 p_3^{10} p_4^5+10654446880 p_1^4 p_2^7 p_3^{11} p_4^5+3277206240 p_1^3 p_2^8 p_3^{11} p_4^5+3557400000 p_1^4 p_2^8 p_3^9 p_4^6+21026246976 p_1^4 p_2^7 p_3^{10} p_4^6+2414513024 p_1^3 p_2^8 p_3^{10} p_4^6+2179888480 p_1^3 p_2^7 p_3^{11} p_4^6+43827168 p_1^3 p_2^7 p_3^{10} p_4^7) \textcolor{red}{z^{27}} 
+(651974400 p_1^5 p_2^9 p_3^{10} p_4^4+596090880 p_1^5 p_2^8 p_3^{11} p_4^4+2151515520 p_1^4 p_2^9 p_3^{11} p_4^4+446054400 p_1^4 p_2^8 p_3^{12} p_4^4+3451486500 p_1^5 p_2^8 p_3^{10} p_4^5+7171718400 p_1^4 p_2^9 p_3^{10} p_4^5+15327479700 p_1^4 p_2^8 p_3^{11} p_4^5+234788400 p_1^3 p_2^8 p_3^{12} p_4^5+499167900 p_1^5 p_2^7 p_3^{10} p_4^6+15028648200 p_1^4 p_2^8 p_3^{10} p_4^6+5685080940 p_1^4 p_2^7 p_3^{11} p_4^6+1285029900 p_1^3 p_2^8 p_3^{11} p_4^6+136959900 p_1^3 p_2^7 p_3^{12} p_4^6+112058100 p_1^4 p_2^7 p_3^{10} p_4^7+82175940 p_1^3 p_2^7 p_3^{11} p_4^7) \textcolor{red}{z^{28}} 
+(391184640 p_1^5 p_2^9 p_3^{11} p_4^4+292723200 p_1^4 p_2^9 p_3^{12} p_4^4+1303948800 p_1^5 p_2^9 p_3^{10} p_4^5+1540871640 p_1^5 p_2^8 p_3^{11} p_4^5+5439772800 p_1^4 p_2^9 p_3^{11} p_4^5+1611502200 p_1^4 p_2^8 p_3^{12} p_4^5+1189465200 p_1^5 p_2^8 p_3^{10} p_4^6+3585859200 p_1^4 p_2^9 p_3^{10} p_4^6+9178974000 p_1^4 p_2^8 p_3^{11} p_4^6+398428800 p_1^4 p_2^7 p_3^{12} p_4^6+180457200 p_1^3 p_2^8 p_3^{12} p_4^6+96049800 p_1^4 p_2^8 p_3^{10} p_4^7+239057280 p_1^4 p_2^7 p_3^{11} p_4^7+70436520 p_1^3 p_2^8 p_3^{11} p_4^7) \textcolor{red}{z^{29}} 
+(60555264 p_1^5 p_2^9 p_3^{12} p_4^4+1201272380 p_1^5 p_2^9 p_3^{11} p_4^5+220774400 p_1^5 p_2^8 p_3^{12} p_4^5+1396604748 p_1^4 p_2^9 p_3^{12} p_4^5+707437500 p_1^5 p_2^9 p_3^{10} p_4^6+816487980 p_1^5 p_2^8 p_3^{11} p_4^6+4106272940 p_1^4 p_2^9 p_3^{11} p_4^6+2005022376 p_1^4 p_2^8 p_3^{12} p_4^6+5478396 p_1^3 p_2^9 p_3^{12} p_4^6+525427980 p_1^4 p_2^8 p_3^{11} p_4^7+12782924 p_1^3 p_2^8 p_3^{12} p_4^7) \textcolor{red}{z^{30}} 
+(435558816 p_1^5 p_2^9 p_3^{12} p_4^5+121968000 p_1^4 p_2^9 p_3^{13} p_4^5+984060000 p_1^5 p_2^9 p_3^{11} p_4^6+191866752 p_1^5 p_2^8 p_3^{12} p_4^6+1896293952 p_1^4 p_2^9 p_3^{12} p_4^6+142296000 p_1^4 p_2^8 p_3^{13} p_4^6+58212000 p_1^5 p_2^8 p_3^{11} p_4^7+274428000 p_1^4 p_2^9 p_3^{11} p_4^7+175877856 p_1^4 p_2^8 p_3^{12} p_4^7) \textcolor{red}{z^{31}} 
+(11176704 p_1^5 p_2^{10} p_3^{12} p_4^5+56770560 p_1^5 p_2^9 p_3^{13} p_4^5+598635576 p_1^5 p_2^9 p_3^{12} p_4^6+30735936 p_1^4 p_2^{10} p_3^{12} p_4^6+420637140 p_1^4 p_2^9 p_3^{13} p_4^6+77962500 p_1^5 p_2^9 p_3^{11} p_4^7+23769900 p_1^5 p_2^8 p_3^{12} p_4^7+231708708 p_1^4 p_2^9 p_3^{12} p_4^7+17431260 p_1^4 p_2^8 p_3^{13} p_4^7+2614689 p_1^4 p_2^8 p_3^{12} p_4^8) \textcolor{red}{z^{32}} 
+(4139520 p_1^5 p_2^{10} p_3^{13} p_4^5+205800 p_1^6 p_2^9 p_3^{12} p_4^6+24147200 p_1^5 p_2^{10} p_3^{12} p_4^6+171990280 p_1^5 p_2^9 p_3^{13} p_4^6+22767360 p_1^4 p_2^{10} p_3^{13} p_4^6+41164200 p_1^4 p_2^9 p_3^{14} p_4^6+93601200 p_1^5 p_2^9 p_3^{12} p_4^7+81650800 p_1^4 p_2^9 p_3^{13} p_4^7+6225450 p_1^4 p_2^9 p_3^{12} p_4^8) \textcolor{red}{z^{33}} 
+(396900 p_1^6 p_2^{10} p_3^{12} p_4^6+18918900 p_1^5 p_2^{10} p_3^{13} p_4^6+23654400 p_1^5 p_2^9 p_3^{14} p_4^6+5336100 p_1^4 p_2^{10} p_3^{14} p_4^6+4365900 p_1^5 p_2^{10} p_3^{12} p_4^7+43423380 p_1^5 p_2^9 p_3^{13} p_4^7+3201660 p_1^4 p_2^{10} p_3^{13} p_4^7+12450900 p_1^4 p_2^9 p_3^{14} p_4^7+2546775 p_1^5 p_2^9 p_3^{12} p_4^8+3735270 p_1^4 p_2^9 p_3^{13} p_4^8) \textcolor{red}{z^{34}} 
+(423360 p_1^6 p_2^{10} p_3^{13} p_4^6+6044544 p_1^5 p_2^{10} p_3^{14} p_4^6+6035040 p_1^5 p_2^{10} p_3^{13} p_4^7+9055200 p_1^5 p_2^9 p_3^{14} p_4^7+1707552 p_1^4 p_2^{10} p_3^{14} p_4^7+2716560 p_1^5 p_2^9 p_3^{13} p_4^8+996072 p_1^4 p_2^9 p_3^{14} p_4^8) \textcolor{red}{z^{35}} 
+(200704 p_1^6 p_2^{10} p_3^{14} p_4^6+369600 p_1^5 p_2^{11} p_3^{14} p_4^6+73500 p_1^6 p_2^{10} p_3^{13} p_4^7+3132668 p_1^5 p_2^{10} p_3^{14} p_4^7+404250 p_1^5 p_2^{10} p_3^{13} p_4^8+916839 p_1^5 p_2^9 p_3^{14} p_4^8+148225 p_1^4 p_2^{10} p_3^{14} p_4^8) \textcolor{red}{z^{36}} 
+(24192 p_1^6 p_2^{11} p_3^{14} p_4^6+88200 p_1^6 p_2^{10} p_3^{14} p_4^7+266112 p_1^5 p_2^{11} p_3^{14} p_4^7+64680 p_1^5 p_2^{10} p_3^{15} p_4^7+407484 p_1^5 p_2^{10} p_3^{14} p_4^8) \textcolor{red}{z^{37}} 
+(18900 p_1^6 p_2^{11} p_3^{14} p_4^7+13860 p_1^5 p_2^{11} p_3^{15} p_4^7+11025 p_1^6 p_2^{10} p_3^{14} p_4^8+51975 p_1^5 p_2^{11} p_3^{14} p_4^8+16170 p_1^5 p_2^{10} p_3^{15} p_4^8) \textcolor{red}{z^{38}} 
+(1120 p_1^6 p_2^{11} p_3^{15} p_4^7+4200 p_1^6 p_2^{11} p_3^{14} p_4^8+6160 p_1^5 p_2^{11} p_3^{15} p_4^8) \textcolor{red}{z^{39}} 
+(630 p_1^6 p_2^{11} p_3^{15} p_4^8+231 p_1^5 p_2^{11} p_3^{16} p_4^8) \textcolor{red}{z^{40}} +42 p_1^6 p_2^{11} p_3^{16} p_4^8 \textcolor{red}{z^{41}}+p_1^6 p_2^{12} p_3^{16} p_4^8 \textcolor{red}{z^{42}},
\)
\begin{equation}\label{H2}\end{equation}
\item[$H_3=$]\(
1+30 p_3 \textcolor{red}{z}+(315 p_2 p_3+120 p_3 p_4) \textcolor{red}{z^2}+(770 p_1 p_2 p_3+1050 p_2 p_3^2+2240 p_2 p_3 p_4) \textcolor{red}{z^3} 
+(5775 p_1 p_2 p_3^2+6930 p_1 p_2 p_3 p_4+14700 p_2 p_3^2 p_4) \textcolor{red}{z^4}+(9702 p_1 p_2^2 p_3^2+90552 p_1 p_2 p_3^2 p_4+31500 p_2 p_3^3 p_4+10752 p_2 p_3^2 p_4^2) \textcolor{red}{z^5} 
+(8085 p_1 p_2^2 p_3^3+161700 p_1 p_2^2 p_3^2 p_4+249480 p_1 p_2 p_3^3 p_4+36750 p_2^2 p_3^3 p_4+92400 p_1 p_2 p_3^2 p_4^2+45360 p_2 p_3^3 p_4^2) \textcolor{red}{z^6} 
+(1181400 p_1 p_2^2 p_3^3 p_4+316800 p_1 p_2^2 p_3^2 p_4^2+443520 p_1 p_2 p_3^3 p_4^2+94080 p_2^2 p_3^3 p_4^2) \textcolor{red}{z^7} 
+(177870 p_1^2 p_2^2 p_3^3 p_4+1358280 p_1 p_2^3 p_3^3 p_4+782100 p_1 p_2^2 p_3^4 p_4+3490575 p_1 p_2^2 p_3^3 p_4^2+44100 p_2^2 p_3^4 p_4^2) \textcolor{red}{z^8} 
+(830060 p_1^2 p_2^3 p_3^3 p_4+2633400 p_1 p_2^3 p_3^4 p_4+711480 p_1^2 p_2^2 p_3^3 p_4^2+4928000 p_1 p_2^3 p_3^3 p_4^2+5035250 p_1 p_2^2 p_3^4 p_4^2+168960 p_1 p_2^2 p_3^3 p_4^3) \textcolor{red}{z^9} 
+(2144604 p_1^2 p_2^3 p_3^4 p_4+1559250 p_1 p_2^3 p_3^5 p_4+3811500 p_1^2 p_2^3 p_3^3 p_4^2+853776 p_1^2 p_2^2 p_3^4 p_4^2+16967181 p_1 p_2^3 p_3^4 p_4^2 
+3234000 p_1 p_2^2 p_3^5 p_4^2+1474704 p_1 p_2^2 p_3^4 p_4^3) \textcolor{red}{z^{10}} 
+(2439360 p_1^2 p_2^3 p_3^5 p_4+18117750 p_1^2 p_2^3 p_3^4 p_4^2+26826030 p_1 p_2^3 p_3^5 p_4^2 +5174400 p_1 p_2^3 p_3^4 p_4^3+2069760 p_1 p_2^2 p_3^5 p_4^3) \textcolor{red}{z^{11}} 
+(711480 p_1^2 p_2^4 p_3^5 p_4+2371600 p_1^2 p_2^4 p_3^4 p_4^2+38368225 p_1^2 p_2^3 p_3^5 p_4^2+6338640 p_1 p_2^4 p_3^5 p_4^2 
+14437500 p_1 p_2^3 p_3^6 p_4^2+5336100 p_1^2 p_2^3 p_3^4 p_4^3+18929680 p_1 p_2^3 p_3^5 p_4^3) \textcolor{red}{z^{12}} 
+(21783930 p_1^2 p_2^4 p_3^5 p_4^2+32524800 p_1^2 p_2^3 p_3^6 p_4^2+8731800 p_1 p_2^4 p_3^6 p_4^2+29988000 p_1^2 p_2^3 p_3^5 p_4^3+8279040 p_1 p_2^4 p_3^5 p_4^3 
+16678200 p_1 p_2^3 p_3^6 p_4^3+1774080 p_1 p_2^3 p_3^5 p_4^4) \textcolor{red}{z^{13}} 
+(1584660 p_1^3 p_2^4 p_3^5 p_4^2+46973475 p_1^2 p_2^4 p_3^6 p_4^2+25194480 p_1^2 p_2^4 p_3^5 p_4^3+43705200 p_1^2 p_2^3 p_3^6 p_4^3 
+17948700 p_1 p_2^4 p_3^6 p_4^3+5488560 p_1^2 p_2^3 p_3^5 p_4^4+4527600 p_1 p_2^3 p_3^6 p_4^4) \textcolor{red}{z^{14}} 
+(5588352 p_1^3 p_2^4 p_3^6 p_4^2+15937152 p_1^2 p_2^5 p_3^6 p_4^2+5808000 p_1^2 p_2^4 p_3^7 p_4^2+3234000 p_1^3 p_2^4 p_3^5 p_4^3+93982512 p_1^2 p_2^4 p_3^6 p_4^3 +3234000 p_1 p_2^4 p_3^7 p_4^3+5808000 p_1^2 p_2^4 p_3^5 p_4^4+15937152 p_1^2 p_2^3 p_3^6 p_4^4+5588352 p_1 p_2^4 p_3^6 p_4^4) \textcolor{red}{z^{15}} 
+(4527600 p_1^3 p_2^5 p_3^6 p_4^2+5488560 p_1^2 p_2^5 p_3^7 p_4^2+17948700 p_1^3 p_2^4 p_3^6 p_4^3+43705200 p_1^2 p_2^5 p_3^6 p_4^3 +25194480 p_1^2 p_2^4 p_3^7 p_4^3+46973475 p_1^2 p_2^4 p_3^6 p_4^4+1584660 p_1 p_2^4 p_3^7 p_4^4) \textcolor{red}{z^{16}} 
+(1774080 p_1^3 p_2^5 p_3^7 p_4^2+16678200 p_1^3 p_2^5 p_3^6 p_4^3+8279040 p_1^3 p_2^4 p_3^7 p_4^3+29988000 p_1^2 p_2^5 p_3^7 p_4^3 +8731800 p_1^3 p_2^4 p_3^6 p_4^4+32524800 p_1^2 p_2^5 p_3^6 p_4^4+21783930 p_1^2 p_2^4 p_3^7 p_4^4) \textcolor{red}{z^{17}} 
+(18929680 p_1^3 p_2^5 p_3^7 p_4^3+5336100 p_1^2 p_2^5 p_3^8 p_4^3+14437500 p_1^3 p_2^5 p_3^6 p_4^4+6338640 p_1^3 p_2^4 p_3^7 p_4^4 +38368225 p_1^2 p_2^5 p_3^7 p_4^4+2371600 p_1^2 p_2^4 p_3^8 p_4^4+711480 p_1^2 p_2^4 p_3^7 p_4^5) \textcolor{red}{z^{18}} 
+(2069760 p_1^3 p_2^6 p_3^7 p_4^3+5174400 p_1^3 p_2^5 p_3^8 p_4^3+26826030 p_1^3 p_2^5 p_3^7 p_4^4+18117750 p_1^2 p_2^5 p_3^8 p_4^4+2439360 p_1^2 p_2^5 p_3^7 p_4^5) \textcolor{red}{z^{19}} 
+(1474704 p_1^3 p_2^6 p_3^8 p_4^3+3234000 p_1^3 p_2^6 p_3^7 p_4^4+16967181 p_1^3 p_2^5 p_3^8 p_4^4+853776 p_1^2 p_2^6 p_3^8 p_4^4 +3811500 p_1^2 p_2^5 p_3^9 p_4^4+1559250 p_1^3 p_2^5 p_3^7 p_4^5+2144604 p_1^2 p_2^5 p_3^8 p_4^5) \textcolor{red}{z^{20}} 
+(168960 p_1^3 p_2^6 p_3^9 p_4^3+5035250 p_1^3 p_2^6 p_3^8 p_4^4+4928000 p_1^3 p_2^5 p_3^9 p_4^4+711480 p_1^2 p_2^6 p_3^9 p_4^4+2633400 p_1^3 p_2^5 p_3^8 p_4^5+830060 p_1^2 p_2^5 p_3^9 p_4^5) \textcolor{red}{z^{21}} 
+(44100 p_1^4 p_2^6 p_3^8 p_4^4+3490575 p_1^3 p_2^6 p_3^9 p_4^4+782100 p_1^3 p_2^6 p_3^8 p_4^5+1358280 p_1^3 p_2^5 p_3^9 p_4^5+177870 p_1^2 p_2^6 p_3^9 p_4^5) \textcolor{red}{z^{22}} 
+(94080 p_1^4 p_2^6 p_3^9 p_4^4+443520 p_1^3 p_2^7 p_3^9 p_4^4+316800 p_1^3 p_2^6 p_3^{10} p_4^4+1181400 p_1^3 p_2^6 p_3^9 p_4^5) \textcolor{red}{z^{23}} 
+(45360 p_1^4 p_2^7 p_3^9 p_4^4+92400 p_1^3 p_2^7 p_3^{10} p_4^4+36750 p_1^4 p_2^6 p_3^9 p_4^5+249480 p_1^3 p_2^7 p_3^9 p_4^5+161700 p_1^3 p_2^6 p_3^{10} p_4^5+8085 p_1^3 p_2^6 p_3^9 p_4^6) \textcolor{red}{z^{24}} 
+(10752 p_1^4 p_2^7 p_3^{10} p_4^4+31500 p_1^4 p_2^7 p_3^9 p_4^5+90552 p_1^3 p_2^7 p_3^{10} p_4^5+9702 p_1^3 p_2^6 p_3^{10} p_4^6) \textcolor{red}{z^{25}} 
+(14700 p_1^4 p_2^7 p_3^{10} p_4^5+6930 p_1^3 p_2^7 p_3^{11} p_4^5+5775 p_1^3 p_2^7 p_3^{10} p_4^6) \textcolor{red}{z^{26}}
+(2240 p_1^4 p_2^7 p_3^{11} p_4^5+1050 p_1^4 p_2^7 p_3^{10} p_4^6+770 p_1^3 p_2^7 p_3^{11} p_4^6) \textcolor{red}{z^{27}}  
+(120 p_1^4 p_2^8 p_3^{11} p_4^5+315 p_1^4 p_2^7 p_3^{11} p_4^6) \textcolor{red}{z^{28}} 
+30 p_1^4 p_2^8 p_3^{11} p_4^6 \textcolor{red}{z^{29}} +p_1^4 p_2^8 p_3^{12} p_4^6 \textcolor{red}{z^{30}},
\)
\begin{equation}\label{H3}\end{equation}
\item[$H_4=$]\(
1+16 p_4 \textcolor{red}{z}+120 p_3 p_4 \textcolor{red}{z^2}+560 p_2 p_3 p_4 \textcolor{red}{z^3}+(770 p_1 p_2 p_3 p_4+1050 p_2 p_3^2 p_4) \textcolor{red}{z^4}+(3696 p_1 p_2 p_3^2 p_4+672 p_2 p_3^2 p_4^2) \textcolor{red}{z^5} 
+(4312 p_1 p_2^2 p_3^2 p_4+3696 p_1 p_2 p_3^2 p_4^2) \textcolor{red}{z^6}+(2640 p_1 p_2^2 p_3^3 p_4+8800 p_1 p_2^2 p_3^2 p_4^2) \textcolor{red}{z^7}+12870 p_1 p_2^2 p_3^3 p_4^2 \textcolor{red}{z^8} 
+(8800 p_1 p_2^2 p_3^4 p_4^2+2640 p_1 p_2^2 p_3^3 p_4^3) \textcolor{red}{z^9}+(3696 p_1 p_2^3 p_3^4 p_4^2+4312 p_1 p_2^2 p_3^4 p_4^3) \textcolor{red}{z^{10}}+(672 p_1^2 p_2^3 p_3^4 p_4^2+3696 p_1 p_2^3 p_3^4 p_4^3) \textcolor{red}{z^{11}} 
+(1050 p_1^2 p_2^3 p_3^4 p_4^3+770 p_1 p_2^3 p_3^5 p_4^3) \textcolor{red}{z^{12}}+560 p_1^2 p_2^3 p_3^5 p_4^3 \textcolor{red}{z^{13}}+120 p_1^2 p_2^4 p_3^5 p_4^3 \textcolor{red}{z^{14}}+16 p_1^2 p_2^4 p_3^6 p_4^3 \textcolor{red}{z^{15}}+p_1^2 p_2^4 p_3^6 p_4^4 \textcolor{red}{z^{16}}.
\)
\begin{equation}\label{H4}\end{equation}              
\end{itemize}              
Let us denote
      \begin{equation}
        \label{3e.5}
         H_s \equiv H_s(z) = H_s(z, (p_i)), \quad  (p_i) = \vec{p} \equiv (p_1,p_2,p_3,p_4).
       \end{equation}
             
      One can easily write down the asymptotic behaviour of the polynomials obtained:
      \begin{equation}
        \label{3e.6}
         H_s = H_s(z, (p_i) )  \sim \left( \prod_{l=1}^{4} (p_l)^{\nu^{sl}} \right) z^{n_s} \equiv 
         H_s^{as}(z, (p_i)), \quad \text{ as } z \to \infty,
       \end{equation}
     where we introduced the integer valued matrix $\nu = (\nu^{sl})$ having the form 
\begin{align}
         \label{3e.7}
        \nu &=  
         \begin{pmatrix}
          1 & 1 & 1 & 1 \\
          1 & 2 & 2 & 1 \\
          1 & 2 & 2 & 1 \\
          1 & 1 & 1 & 1 \\
         \end{pmatrix},
         \quad  
         \begin{pmatrix}
          2 & 2 & 2 & 2 \\
          2 & 4 & 4 & 4 \\
          2 & 4 & 6 & 6 \\
          1 & 2 & 3 & 4 \\
         \end{pmatrix},
         \quad 
         \begin{pmatrix}
          2 & 2 & 2 & 1 \\
          2 & 4 & 4 & 2 \\
          2 & 4 & 6 & 3 \\
          2 & 4 & 6 & 4 \\ 
         \end{pmatrix},
         \quad \nonumber \\
        &\hspace{5em} \begin{pmatrix}
          2 & 2 & 1 & 1 \\
          2 & 4 & 2 & 2 \\
          1 & 2 & 2 & 1 \\
          1 & 2 & 1 & 2 \\ 
         \end{pmatrix},
         \quad 
         \begin{pmatrix}
                   4 & 6 & 8 & 4 \\
                   6 & 12 & 16 & 8 \\
                   4 & 8 & 12 & 6 \\
                   2 & 4 & 6 & 4 \\ 
                  \end{pmatrix}
\end{align}
      for Lie algebras $A_4,  B_4,  C_4, D_4, F_4$, respectively.
      In these five cases there is a simple property 
                   \begin{equation}
                     \label{3e.10a}
                      \sum_{l= 1}^4 \nu^{sl} = n_s, \quad s = 1, 2, 3, 4.
                    \end{equation}
                                       
     Note that for Lie algebras  $B_4$, $C_4$, $D_4$ and $F_4$ we have 
      \begin{equation}
        \label{3e.8a}
      \nu({\cal G}) = 2 A^{-1}, \quad {\cal G}=B_4, C_4, D_4, F_4,
      \end{equation}
      where $A^{-1}$ is inverse Cartan matrix, whereas in the $A_4$-case   
      the matrix $\nu$ is related to the inverse Cartan matrix as follows
      \begin{equation}
        \label{3e.8b}
         \nu({\cal G})  = A^{-1} (I + P), \quad {\cal G}=A_4.
       \end{equation}
       Here $I$ is $4 \times 4$ identity matrix and
      \begin{equation}
         \label{3e.9}
         P  = 
         \begin{pmatrix}
          0 & 0 & 0 & 1 \\
          0 & 0 & 1 & 0 \\
          0 & 1 & 0 & 0 \\
          1 & 0 & 0 & 0 \\
          \end{pmatrix}
         \end{equation}
      is a permutation matrix corresponding to the permutation $\sigma \in S_4$ ($S_4$ is symmetric group)
      \begin{equation}
        \label{3.10}
         \sigma: (1,2,3,4) \mapsto (4,3,2,1),
       \end{equation}
       by the following relation  $P = (P^i_j) = (\delta^i_{\sigma(j)})$. 
       Here $\sigma$   is the generator of the group 
       $G = \{ \sigma, {\rm id} \}$, which is the group of symmetry of the Dynkin diagram for $A_4$. $G$ is 
        isomorphic to the group $\mathbb{Z}_2$.

In case of $D_4$ the  group of symmetry of the Dynkin diagram  $G'$ is isomorphic to the symmetric group $S_3$ acting on the set of three vertices $\{1, 3, 4 \}$ of the Dynkin diagram via their  permutations. 
    The existence of the above symmetry groups $G \cong \mathbb{Z}_2$ and $G' \cong S_3$  implies certain identity properties for the fluxbrane polynomials $H_s(z)$.
  
         Let us denote $\hat{p}_i = p_{\sigma(i)} $ for the $A_4$ case, and
       $\hat{p}_i = p_{i}$ for $B_4$, $C_4$, $D_4$, $F_4$ cases  ($i= 1,2,3,4$).         
       We call the ordered set $(\hat{p}_i)$ as \textit{dual} one to the ordered set $(p_i)$. It corresponds to the action (trivial or nontrivial) of the group $\mathbb{Z}_2$ on vertices of the Dynkin diagrams for above algebras. 
  
       Then we obtain the following identities which were directly verified by using MATHEMATICA algorithms. 
   \\[1em] 
       {\bf Symmetry relations.}
   \\
       {\bf Proposition 1.} {\em  
         The fluxbrane polynomials obey for all $p_i$ and $z>0$
                 the identities:
        \bearr{3.11}
        H_{\sigma(s)}(z, (p_i) )\, = H_s(z, (\hat{p}_i)) \qquad \text{ for $A_4$ case},\nnn
        H_{\sigma'(s)}(z, (p_i) ) = H_s(z, (p_{\sigma'(i)})\qquad\! \text{ for $D_4$ case},
         \ear
         for any $\sigma' \in S_3$, $s= 1, \dots, 4$.  }
         We call  relations (\ref{3.11})  as symmetry ones.
   \\[1em]
       {\bf Duality relations.}
   \\
       {\bf Proposition 2.} {\em  The fluxbrane polynomials
        corresponding to Lie algebras $A_4$,
        $B_4$, $C_4$, $D_4$ and $F_4$ obey for all $p_i > 0$ and $z > 0$
        the identities
        \begin{equation}
          \label{3.12}
           H_{s}(z, (p_i) ) = H_s^{as}(z, (p_i)) H_s(z^{-1}, (\hat{p}_i^{-1})),
         \end{equation}
         $s = 1, 2, 3, 4$. }

         We call  relations (\ref{3.12})  as duality ones. These relations may be used 
          for deriving  $1/\rho$-expansion for the metric and the fields involved at  large distances. 
 
       {\bf Fluxes.}
       Let us consider an oriented $2$-dimensional manifold 
      $M_{*} =(0, + \infty) \times S^1$, $R > 0$. 
      One can calculate the  flux integrals over this manifold:
       \beq{3.19}
        \Phi^s  = \int_{M_{*}} F^s =
         2 \pi \int_{0}^{ + \infty} d \rho \rho {\cal B}^s ,
       \eeq    
       where 
        \begin{equation}
           \label{3e.16}
           {\cal B}^s =   q_s  \prod_{l = 1}^{4}  H_{l}^{- A_{s l}}.
        \end{equation} 
       \\
       The  flux integrals $\Phi^s$  are convergent and read as follows        
      \cite{Ivas-flux-17}
     \beq{3.25}
       \Phi^s =    4 \pi n_s q_s^{-1} h_s,  
      \eeq
   $s =1, 2,3, 4$. Thus, any flux $\Phi^s$ depends upon one integration constant  $q_s \neq 0$, while the integrand form $F^s$ depends upon all constants: $q_1, q_2, q_3, q_4$.
  
    In the $A_4$-case we have:
      \beq{3.A2f}
     (q_1 \Phi^1, q_2 \Phi^2, q_3 \Phi^3, q_4 \Phi^4) =   4 \pi h  (4,6,6,4),
      \eeq
      where $h_1 = h_2 = h_3 = h_4 = h$. 
     
      In the $B_4$=case we find: 
      \beq{3.Bf}
      (q_1 \Phi^1, q_2 \Phi^2, q_3 \Phi^3, q_4 \Phi^4) =  
       4 \pi  ( 8 h_1, 14 h_2, 18 h_3, 10 h_4 ) =
      4 \pi h ( 8, 14, 18, 20),
      \eeq
     where $h_1 =  h_2 = h_3 = h$, $h_4 = 2h $.
  
  In the $C_4$-case we obtain: 
   \beq{3.Cf}
   (q_1 \Phi^1, q_2 \Phi^2, q_3 \Phi^3, q_4 \Phi^4) =   
   4 \pi  ( 7 h_1, 12 h_2, 15 h_3, 16 h_4) =
        4 \pi h ( 7, 12, 15, 8),
    \eeq
   where $h_1 =  h_2 = h_3 = h$, $h_4 = \frac{1}{2} h $.  
 
  In the $D_4$-case we are led to relations: 
    \beq{3.Df}
    (q_1 \Phi^1, q_2 \Phi^2, q_3 \Phi^3, q_4 \Phi^4) =   4 \pi h ( 6, 10, 6, 6),
     \eeq
    where $h_1 =  h_2 = h_3 = h_4 = h$.  (In all examples relations (\ref{3.26h})
    are used.)
  
  In the $F_4$-case we similarly obtain:
   \beq{3.Ff}
         (q_1 \Phi^1, q_2 \Phi^2, q_3 \Phi^3, q_4 \Phi^4) =  
          4 \pi  ( 22 h_1, 42 h_2, 30 h_3, 16 h_4 ) =
         4 \pi h ( 22, 42, 60, 32),
         \eeq
        where $h_1 =  h_2 =  h$, $h_3 = h_4 = 2h$.

    For $D =4$ and $g^2 = -  dt \otimes dt + d x \otimes d x$, $q_s$ coincides with 
  the value of the $x$-component of the $s$-th magnetic field on the axis of symmetry, $s =1,2,3,4$. 
    
     We  note also that  by putting $q_1 = 0$ we get the Melvin-type solutions corresponding 
     to Lie algebras $A_3$,  $B_3$,  $C_3$,  $A_3$ and $C_3$, respectively,  which were analyzed 
     in ref. \cite{BolIvas-R3-18}.
     (The case of the rank 2 Lie algebra $G_2$ \cite{BolIvas-R2-17} may be obtained for 
     the $D_4$ case when     $q_1 = q_3 = q_4$.)  
         The case of non-exceptional Lie algebras of rank 4 was considered earlier 
         in  \cite{Bol-Ivas-R4-18}.
         
         {\bf Special solutions.} Let us put         
          $p_1 = p_2= p_3 = p_4 = p > 0$. We get binomial relations
          \beq{3.27}
                   H_{s}(z) =  H_{s}(z;(p,p,p,p)) = (1 + p z)^{n_s},        
          \eeq
         which certainly satisfy the master equations (\ref{1.1}) with boundary 
         conditions (\ref{1.2}) imposed when parameters $q_s$ obey 
         \beq{3.27a}
           \frac{1}{4} K_s q_s^2 / n_s =  p,        
         \eeq
         $s = 1,2,3,4.$
         
         Relation  (\ref{3.27}) is   satisfied for all polynomials presented above.
          One can also readily check the relations 
         for fluxes in (\ref{3.25}) for the special case $p_1 = p_2= p_3 = p_4 = p$.

    \section{Dilatonic black holes for simple Lie algebras of rank 4}
     
     Relations (constraints) on dilatonic coupling vectors  
     (\ref{2.17}), (\ref{2.18}) appear also for dilatonic black hole solutions which   
     are defined on the manifold
    \beq{5.1}
      M' = (R_0, + \infty) \times (M_0 = S^2) \times (M_1 = \R) \times M_2,
     \eeq
     where $R_0 = 2 \mu > 0$ and $M_2$ is a  Ricci-flat manifold. These solutions on the manifold $M'$ from 
     (\ref{5.1})
     for the model under consideration may be extracted from general black brane solutions from  
     refs. \cite{Ivas-Symmetry-17,I-14,IMp3}. They read: 
     \bear{5.1g}
      g= \Bigl(\prod_{s = 1}^{4} {\bf H}_s^{2 h_s /(D-2)} \Bigr)
      \biggl\{  f^{-1} dR \otimes d R  + R^2 g^0  -
      \Bigl(\prod_{s = 1}^{4} {\bf H}^{-2 h_s} \Bigr) f dt \otimes dt +
        g^2  \biggr\},
     \\  \label{5.2p}
      \exp(\varphi^a)=
      \prod_{s = 1}^{4} {\bf H}^{h_s  \lambda_{s}^a},
     \\  \label{5.2f}
      F^s = - Q_s R^{-2} \left( \prod_{l = 1}^{4}  {\bf H}_{l}^{- A_{s
      l}} \right)  dR \wedge d t,
      \ear
     $s,a = 1,2,3,4$, where $f = 1 - 2\mu R^{-1}$, $g^0$ is the standard 
      metric on $M_0 = S^2$ and $g^2$ is a  Ricci-flat metric of 
      signature $(+, \dots, +)$ on $M_{2}$.  Here $Q_s \neq 0$ are integration constants  
     (charges).
    
     The functions ${\bf H}_s = {\bf H}_s (R) > 0$ obey the master equations
    \beq{5.3}
      R^2 \frac{d}{dR} \left( f \frac{ R^2}{{\bf H}_s} \frac{d}{dR} {\bf H}_s \right) =
       B_s \prod_{l = 1}^{4}  {\bf H}_{l}^{- A_{s l}},
      \eeq
     with  the following boundary conditions on the horizon and at infinity imposed:
     \beq{5.4}
       {\bf H}_{s}(R_0 + 0) =  {\bf H}_{s0} > 0, \qquad {\bf H}_{s}(+ \infty) = 1
     \eeq
     where
     \beq{5.5}
      B_s =  - K_s Q_s^2,
     \eeq
     $s = 1, 2,3,4$. Here relations (\ref{2.16}) are also valid. 
     
       For Lie algebras 
     of rank $4$ the functions ${\bf H}_s$ are polynomials 
     of rank (\ref{3.n})   with respect to $ R^{-1}$. 
        By using approach of ref. \cite{I-14} these polynomials may be 
     obtained (at least for small enough $Q_s$) from fluxbrane polynomials $H_s(z)$ 
     presented in this paper extended to negative values of parameters $p_s$. 
     
      Indeed, let us denote $f =  1 -  2\mu/R$.
     Then the relations (\ref{5.3}) may be rewritten as 
     \beq{5.6}
      \frac{d}{df} \left( \frac{f}{{\bf H}_s}
      \frac{d}{df} {\bf H}_s \right) =  B_s (2 \mu )^{-2}
      \prod_{l =1}^{4}  {\bf H}_{l}^{- A_{s l}}, 
      \eeq
       $s = 1,2,3,4$.
      These relations could be solved (at least for small enough $Q_s$) by using fluxbrane
     polynomials $ H_{s}(f) =  H_{s}(f; \vec{p})$, corresponding
     to $4 \times 4$ Cartan matrix $(A_{s l})$, where $\vec{p} =
     (p_1,p_2,p_3,p_4)$ is the set of  parameters. Here
     we impose the restrictions $p_s \neq 0$ for all $s$.
     
     Due to approach of ref. \cite{I-14} we put 
     \beq{5.7}
       {\bf H}_s =   H_{s}(f;\vec{p})/ H_{s}(1;\vec{p})
     \eeq 
     for  $s = 1,2,3,4$. Then the relations (\ref{5.6}), 
     are satisfied identically if \cite{I-14}
     
     \beq{5.7a}
      n_s p_s  \prod_{l =1}^{4}  (H_{l}(1;\vec{p}))^{- A_{s l}} = B_s /(2 \mu)^2,
      \eeq 
     $s = 1,2,3,4$.
     
     We call the set of parameters $\vec{p} = (p_1,p_2,p_3,p_4)$ ($p_i \neq 0$)
      as proper one if \cite{I-14}
       \beq{5.8}
       H_{s}(f;\vec{p}) > 0
      \eeq
      for all $f \in [0,1]$ and $s = 1,2,3,4$.  In what  follows we consider only proper $\vec{p}$.
      In relations (\ref{5.7a}) we have  $p_s < 0$ and $B_s < 0$ for $s = 1,2,3,4$.
     
     The  boundary conditions  (\ref{5.4}) are valid since due to relation (\ref{5.7})
       \beq{5.9}
       {\bf H}_{s}(2\mu + 0) = 1/ H_{s}(1;\vec{p}) > 0,
       \eeq
      $s = 1,2,3,4$.
     
      Locally, for small enough
      $p_i$ the relation  (\ref{5.7a}) defines one-to-one correspondence between
      the sets of parameters $(p_1,p_2,p_3,p_4)$ and $(Q_1^2, Q_2^2,Q_3^2, Q_4^2 )$ and the 
      set $(p_1,p_2,p_3,p_4)$ is proper. 
     
      Relations (\ref{5.9}) imply the following formula for
       the Hawking temperature  \cite{I-14}
      \beq{5.7T}
      T_H =   \frac{1}{8 \pi \mu}
      \prod_{s = 1}^{4} (H_{s}(1;\vec{p}))^{h_s}.  \eeq
     
     {\bf Special solutions. }
     For any algebra under consideration there
     exists a special solution with binomial relations for moduli functions 
     \beq{5.9b}
       {\bf H}_{s} = (1 + P/R)^{n_s},
     \eeq
      with $P > 0$,   if
       \beq{5.9a}
       K_s Q_s^2/n_s = P (P + 2 \mu),
       \eeq
       $s = 1,2,3,4$. This  may be readily verified  by substituting 
       these functions into the  master equations (\ref{5.3}).
       The corresponding fluxbrane polynomials (\ref{3.27}) have coinciding (negative)
      parameters $p_1 = p_2= p_3 = p_4 = p < 0$ which obey
       \beq{5.10}
          - \frac{p}{1 + p} = P/(2\mu) > 0,
        \eeq  
     where $ -1 < p < 0$. (For this values the set $(p,p,p,p)$ is proper one.) 
     Relation (\ref{5.10})  may be extracted just  from  (\ref{5.7}).
       The Hawking temperature  in this case reads as 
       \beq{5.7TT}
           T_H =   \frac{1}{8 \pi \mu} \left(1 + p \right)^{ A} = 
            \frac{1}{8 \pi \mu} \left(1 + \frac{P}{2 \mu}\right)^{- A}, \qquad 
            A = \sum_{s = 1}^{4}  n_s h_s.
        \eeq
    Here the identity $1/(1 + p) = 1 + P/(2\mu)$ is used.  
    
    {\bf Phantom black holes.} 
    Now we consider the case of special solution with $p > 0$. We get from (\ref{5.10})
    $- 2\mu < P < 0$ and due to relation (\ref{5.9a}) we find  $K_s Q_s^2 < 0$ which imply 
    (due to $K_s > 0$)  $Q_s^2 < 0$, i.e. we are led to pure imaginary charges $Q_s$. 
    But one can overcome this point by considering from the very beginning ``phantom'' 
    fields of forms $F^s$, i.e. one should consider the action with wrong signs
    of electromagnetic-type terms
    \beq{5.11}
     S_f =\int d^Dx \sqrt{|g|} \biggl \{R[g]-
     \delta_{a b} g^{MN}\partial_M \varphi^{a} \partial_N \varphi^{b} + \frac{1}{2}
     \sum_{s =1}^{4}\exp[2 \vec{\lambda}_s \vec{\varphi}](F^s)^2 \biggr \},
     \eeq
      instead of (\ref{2.1}). Models with phantom ``electromagnetic-type'' field
       were considered in literature, see for example  \cite{CFR,ACFR}.
       In this case one should replace the relation  (\ref{5.5})  by 
       \beq{5.12}
            B_s =   K_s Q_s^2.
       \eeq  
       For special phantom black hole solutions we obtain
        \beq{5.13}
              - K_s Q_s^2/n_s = P (P + 2 \mu),
        \eeq
       ($- 2\mu < P < 0$) instead of (\ref{5.9a}). In general case the phantom black hole solutions are
       described by formulae (of this Section) presented above with the relation (\ref{5.12}) instead of (\ref{5.5}). These solutions use fluxbrane polynomials with positive $p_i$ which were studied in previous 
       sections.

  \section{\bf Conclusions}
  
    In this paper, the generalized multidimensional 
    family of Melvin-type solutions was considered corresponding to  finite-dimensional Lie algebras of rank $4$:
    ${\cal G} = A_4,  B_4,  C_4, D_4,  F_4$.  Each solution of that family is governed by a set of $4$ fluxbrane polynomials $H_s(z)$, $s =1,2,3,4$. These so-called fluxbrane polynomials define special solutions to open Toda chain equations corresponding      to the Lie algebra  ${\cal G}$.
         
    The polynomials $H_s(z)$ depend also upon parameters $q_s$, which coincides for $D =4$  (up to a sign) with the values  of  colored magnetic fields on the axis of symmetry. 
     
    We have presented the symmetry relations and the duality identities for polynomials 
    under consideration. These identities may be used  
   in deriving  $1/\rho$-expansion for solutions at large distances $\rho$.               
   We have also presented two-dimensional flux integrals $\Phi^s = \int_{M_{*}} F^s$ ($s =1, 2,3,4$) over a 
   two dimensional submanifold $M_{*}$. Each total flux  $\Phi^s$  depends only upon one corresponding parameter $q_s$, whereas the integrand  $F^s$ depends on all parameters $q_s$, $s =1,2,3,4$.

   Here we have  suggested a possible applications of the fluxbrane polynomials under consideration 
    to a class of dilatonic black hole solutions which are analogs of the  Melvin-type solutions. A subclass of special charged  black hole solutions governed by two parameters:
   $P > 0$ and $\mu > 0$, was considered.   It was pointed out that the consideration of  black hole solution 
   in the model with ``phantom'' fields of forms will use original fluxbrane polynomials $ H_{s}(f;(p_1,p_2,p_3,p_4))$, i.e. those which have positive values of parameters $p_i$, $i = 1,2,3,4$. 
   (For usual charged black holes one should deal with negative $p_i$.)
   The detailed consideration of such phantom black holes governed by fluxbrane polynomials (for these and other 
   Lie algebras)  will be a subject of separate paper.
                 
 
    \begin{center}
    {\bf Acknowledgments}
    \end{center}
  
  This paper has been supported by the RUDN University Strategic Academic Leadership Program (recipients: V.D.I. - mathematical model development and S.V.B. - simulation model development).  The reported study was funded by RFBR, project number 19-02-00346 (recipients S.V.B and V.D.I. - physical model development).

 \end{document}